\documentclass[aps,prd,epsf,showpacs,superscriptaddress]{revtex4}
\usepackage{epsfig}
\usepackage{graphicx}
\usepackage{amsmath,amssymb,latexsym}
\usepackage{bm}

\def\lbldef#1#2{\expandafter\gdef\csname #1\endcsname {#2}}

\def\href#1#2{#2}

\newcommand{\bwide}{\begin{widetext}}
\newcommand{\ewide}{\end{widetext}}
\newcommand{\beq}[1]{\begin{equation} \label{(#1)}}
\newcommand{\eeq}{\end{equation}}
\newcommand{\ba}[1]{\begin{eqnarray} \label{(#1)}}
\newcommand{\ea}{\end{eqnarray}}

\begin{document}
\hspace*{130mm}{\large \tt FERMILAB-PUB-07-454-A}

\title{Predictions for the Cosmogenic Neutrino Flux in Light of New
  Data from the Pierre Auger Observatory}

\author{Luis A.~Anchordoqui}
\affiliation{Department of Physics, University of Wisconsin-Milwaukee, 
             P O Box 413, Milwaukee, WI 53201, USA}
\author{Haim Goldberg}
\affiliation{Department of Physics,
Northeastern University, Boston, MA 02115, USA}
\author{Dan Hooper}
\affiliation{Fermi National Accelerator Laboratory, 
             Theoretical Astrophysics, Batavia, IL 60510, USA} 
\author{Subir Sarkar}
\affiliation{Rudolf Peierls Centre for Theoretical Physics, 
             University of Oxford, Oxford OX1 3NP, UK}
\author{Andrew  Taylor}
\affiliation{Max-Planck-Institut f\"ur Kernphysik, 
             Postfach 103980, D-69029 Heidelberg, GERMANY}
\begin{abstract}
  The Pierre Auger Observatory (PAO) has measured the spectrum and
  composition of the ultrahigh energy cosmic rays with unprecedented
  precision. We use these measurements to constrain their spectrum and
  composition as injected from their sources and, in turn, use these
  results to estimate the spectrum of cosmogenic neutrinos generated
  in their propagation through intergalactic space. We find that the
  PAO spectrum and elongation rate measurements can be well fitted if
  the injected cosmic rays consist entirely of nuclei with masses in
  the intermediate (carbon, nitrogen or oxygen) to heavy (iron,
  silicon) range. A mixture of protons and heavier species is also
  acceptable but (on the basis of existing hadronic interaction models) injection of pure light nuclei (protons, helium) results in
  unacceptable fits to the new elongation rate data. The expected
  spectrum of cosmogenic neutrinos can vary considerably, depending on
  the precise spectrum and chemical composition injected from the
  cosmic ray sources.  In the models where heavy nuclei dominate the
  cosmic ray spectrum and few dissociated protons exceed GZK energies,
  the cosmogenic neutrino flux can be suppressed by up to two orders
  of magnitude relative to the all-proton prediction, making its detection
  beyond the reach of current and planned neutrino telescopes.
  Other models consistent with the data, however, are
  proton-dominated with only a small (1-10\%) admixture of heavy
  nuclei and predict an associated cosmogenic flux within
  the reach of upcoming experiments. Thus a detection or non-detection of
  cosmogenic neutrinos can assist in discriminating between these
  possibilities.
\end{abstract}

\pacs{98.70.Sa, 95.85.Ry}

\maketitle

\section{Introduction}

The Pierre Auger Observatory (PAO)~\cite{Abraham:2004dt} has been
designed to study ultrahigh energy cosmic rays (UHECRs) with energies
above about $10^{18}~{\rm eV}$, with the aim of uncovering their
origin and nature. Such events are too rare to be detected directly,
but the direction, energy and (to some extent) the chemical
composition of primary particles can be inferred from the cascade of
secondary particles induced when the primary impinges upon the upper
atmosphere.  These cascades, or air showers, have been studied by
measuring the fluorescence light they produce in the atmosphere and by
directly sampling shower particles at ground level. The PAO is a hybrid
detector, exploiting both of these well established techniques by
employing an array of water \v{C}erenkov detectors overlooked by
fluorescence telescopes. On clear and dark nights, air showers are
simultaneously observed by both types of detectors, facilitating
powerful reconstruction methods and control of the systematic errors
which have plagued previous cosmic ray experiments.

The chemical composition of the UHECRs has long been a subject of
debate. On the one hand, there are both theoretical and observational
motivations for favoring a cosmic ray spectrum dominated by heavy or
intermediate mass nuclei at the highest energies. In particular,
according to the Hillas criterion~\cite{hillas}, plausible
astrophysical sources are able to accelerate particles to a maximum
energy proportional to their charge. It is, therefore, rather less
challenging for cosmic ray accelerators to generate $\sim$$10^{20}$ eV
iron nuclei than protons, for example. Furthermore, the lack of
observed point sources suggests that there are either a very large
number of faint cosmic ray sources, or that these particles are
significantly deflected by large-scale magnetic fields. Since this
deflection is most efficient for particles with large atomic number
(electric charge), nuclei are again favored. On the other hand, it has
been argued that the ``dip'' observed around $\sim 10^{19}$~eV in the
UHECR spectrum is a signature of protons interacting with cosmic
microwave background photons via electron-positron pair
production~\cite{berezinsky}.

To identify the species of primary UHECRs, one has to study the
development of the resulting showers in detail. Essentially, at a
given energy, showers initiated by heavy nuclei develop earlier in the
atmosphere than proton-induced showers. This, however, is complicated
by fluctuations associated with the stochasticity of the first
interaction~\cite{Anchordoqui:2004xb}. On average, proton-induced
showers reach their maximum development deeper in the atmosphere, i.e.
at a larger cumulated grammage, $X_{\rm max}$.  Furthermore, $X_{\rm
  max}$ increases with energy, as more energetic showers can develop
longer before being quenched by atmospheric losses.  The way the
average depth of maximum, $\langle X_{\rm max} \rangle$, varies with
energy depends on the primary composition and particle interactions
according to $\langle X_{\rm max} \rangle = D_e \ln (E/E_0),$ where
$D_e$ is the elongation rate and $E_0$ is a characteristic energy that
depends on the primary composition~\cite{Linsley:gh}. Since $\langle
X_{\rm max} \rangle$ and $D_e$ can be determined from the longitudinal
shower profile as measured with a fluorescence detector, $E_0$ and
thus the composition can be extracted after estimating the energy
from the total fluorescence yield.

The latest results from the PAO were presented recently at the 30th
International Cosmic Ray
Conference~\cite{Roth:2007in,Unger:2007mc,Anchordoqui:2007tb,Bigas:2007tp}.
Even though the observatory will not be completed until the end of
2007, it has already become the preeminent source of UHECR data.  For
example, $X_{\rm max}$ measurements~\cite{Unger:2007mc} at extremely
high energies ($E_{\rm CR} \agt 10^{19.3}~{\rm eV}$) are significantly
more precise (uncertainties smaller by a factor of approximately 4)
than the best measurements from the HiRes
experiment~\cite{Abbasi:2004nz}.

The chemical composition of the highest energy cosmic rays has
important implications to the the field of high and ultrahigh energy
neutrino astronomy. Protons with an energy above a few times $10^{19}$~eV
interact efficiently with the cosmic microwave and infrared background
photons~\cite{gzk}, producing pions which decay to generate a spectrum
of ultrahigh energy neutrinos, known as the cosmogneic neutrino
flux~\cite{cosmogenic}. Cosmogenic neutrinos have often been thought
of as an essentially guaranteed flux of ultrahigh energy neutrinos,
likely within the reach of current and next generation neutrino
detectors such as IceCube~\cite{Ahrens:2002dv} and ANITA~\cite{Barwick:2005hn}, as well as detectors such as the PAO
itself~\cite{augernus}.  This conclusion can be altered, however, if
a substantial fraction of the UHECR spectrum consists of heavy or
intermediate mass nuclei rather than protons~\cite{cosmogenicnuclei}.
Cosmic ray nuclei, in contrast to protons, generate ultrahigh energy
neutrinos through photodisintegration followed by pion production
through nucleon-photon scattering. Depending on the choice of chemical
composition and injected spectrum of the UHECRs, the cosmogenic
neutrino spectrum can in some cases be considerably suppressed
relative to that predicted for an all-proton composition.

In this paper, we consider the recent spectrum and elongation rate
measurements from the PAO and use these results to constrain the
spectrum and chemical composition of UHECRs at their sources.  We then
turn our attention to the cosmogenic neutrino spectrum which is
generated through the interaction of these particles with the cosmic
microwave and infrared backgrounds. We find UHECR injection models
with a wide range of chemical compositions consistent with the
spectrum and elongation rate measurements of Auger. An
all-intermediate mass to all-heavy nuclei composition consistent with
the data can lead to a considerable suppression (up to two orders of
magnitude) of the cosmogenic neutrino flux in comparison to the
all-proton case. However, the data is also consistent with a
proton-dominated spectrum with a small admixture of heavy nuclei, in
which case the cosmogenic neutrino spectrum is very similar to that predicted in the all-proton scenario. In this latter case, kilometer-scale neutrino
telescopes will be expected to observe on the order of one cosmogenic
neutrino event per year. In the former case, the rate will be much
lower and is unlikely to be observed in current or planned
experiments.

\section{The Spectrum and Chemical Composition of Ultrahigh Energy
  Cosmic Rays}

\begin{figure}[!tbp]
\epsfig{file=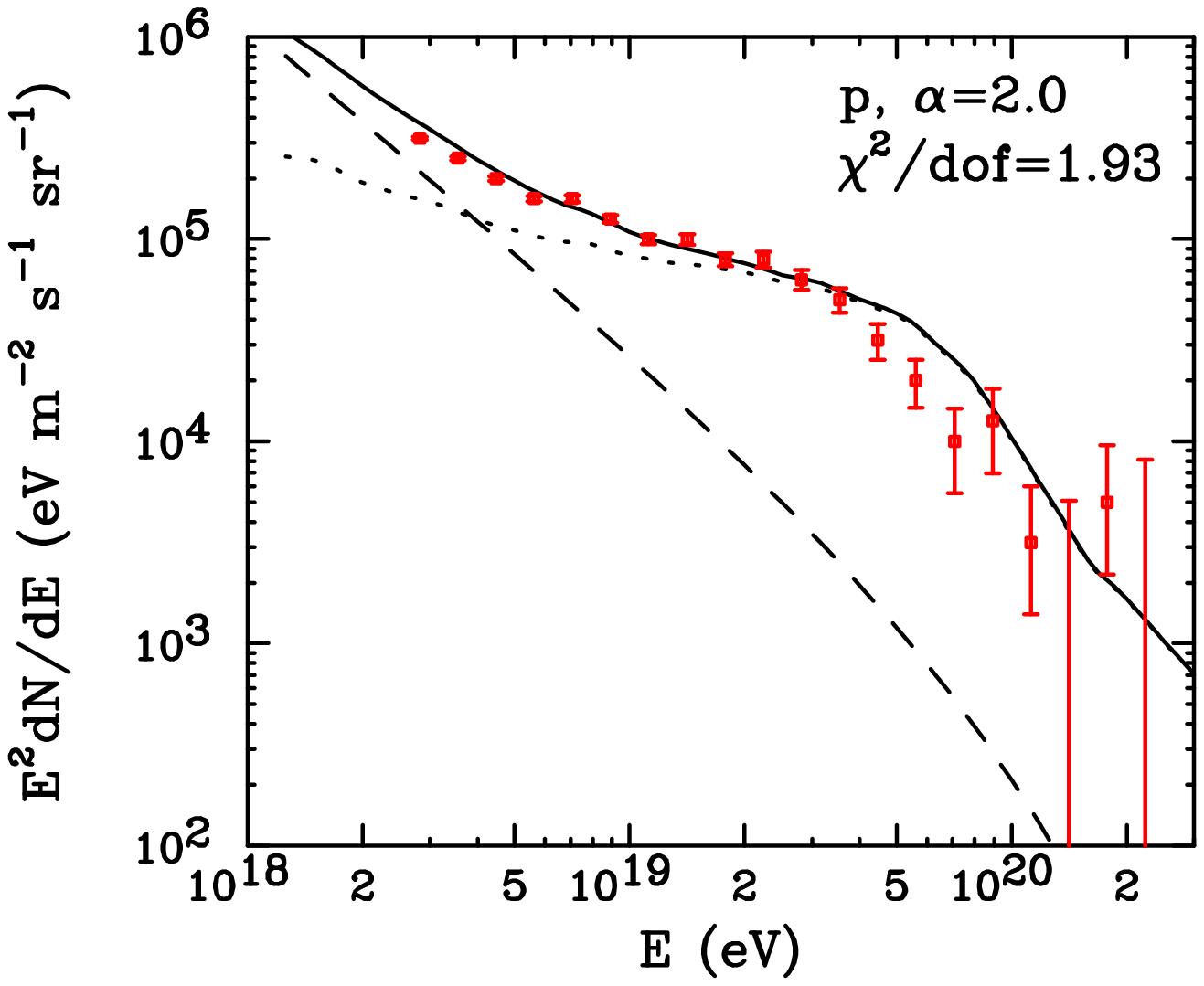,width=.35\columnwidth}
\hspace{1.0cm}
\epsfig{file=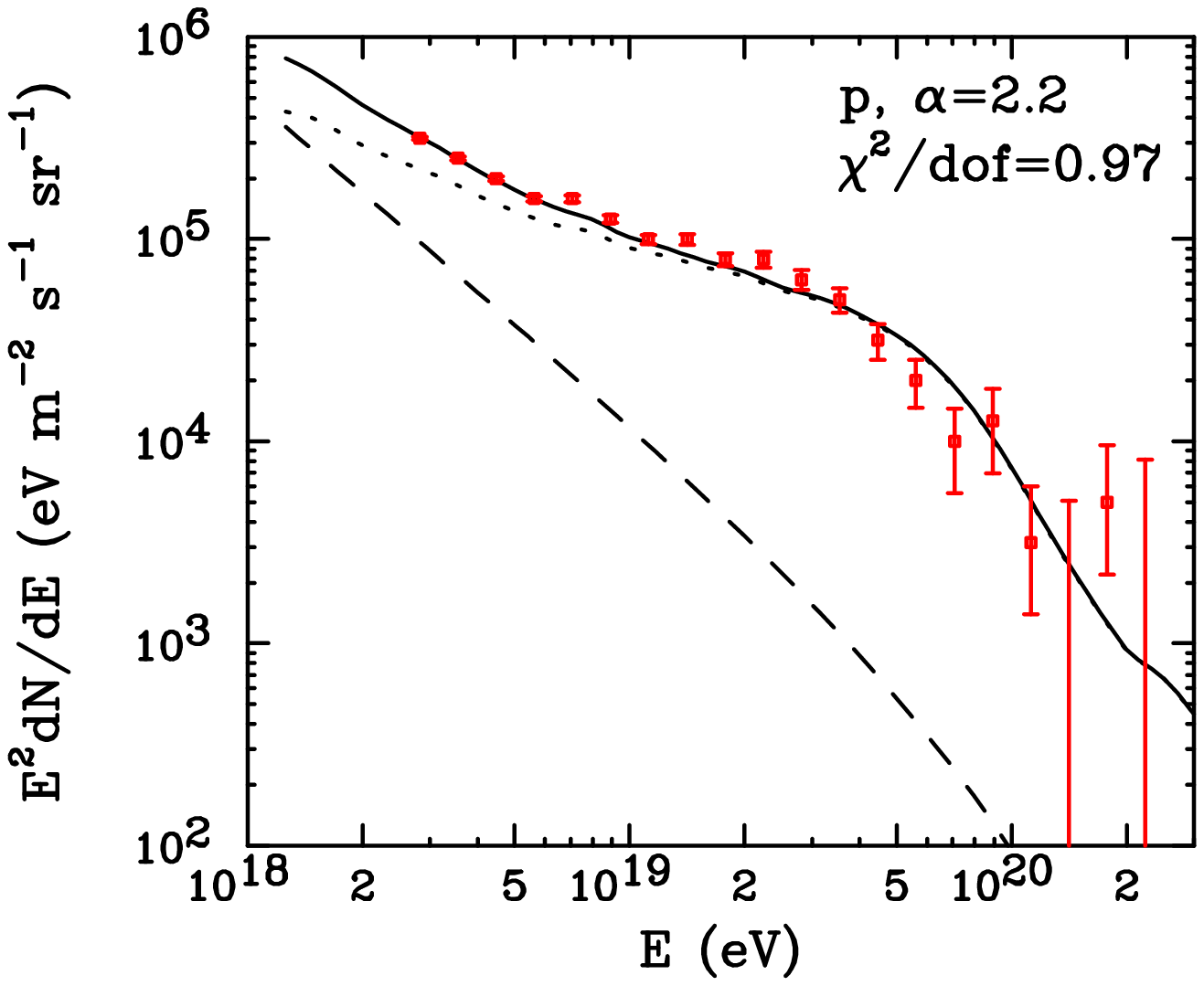,width=.35\columnwidth} \\
\epsfig{file=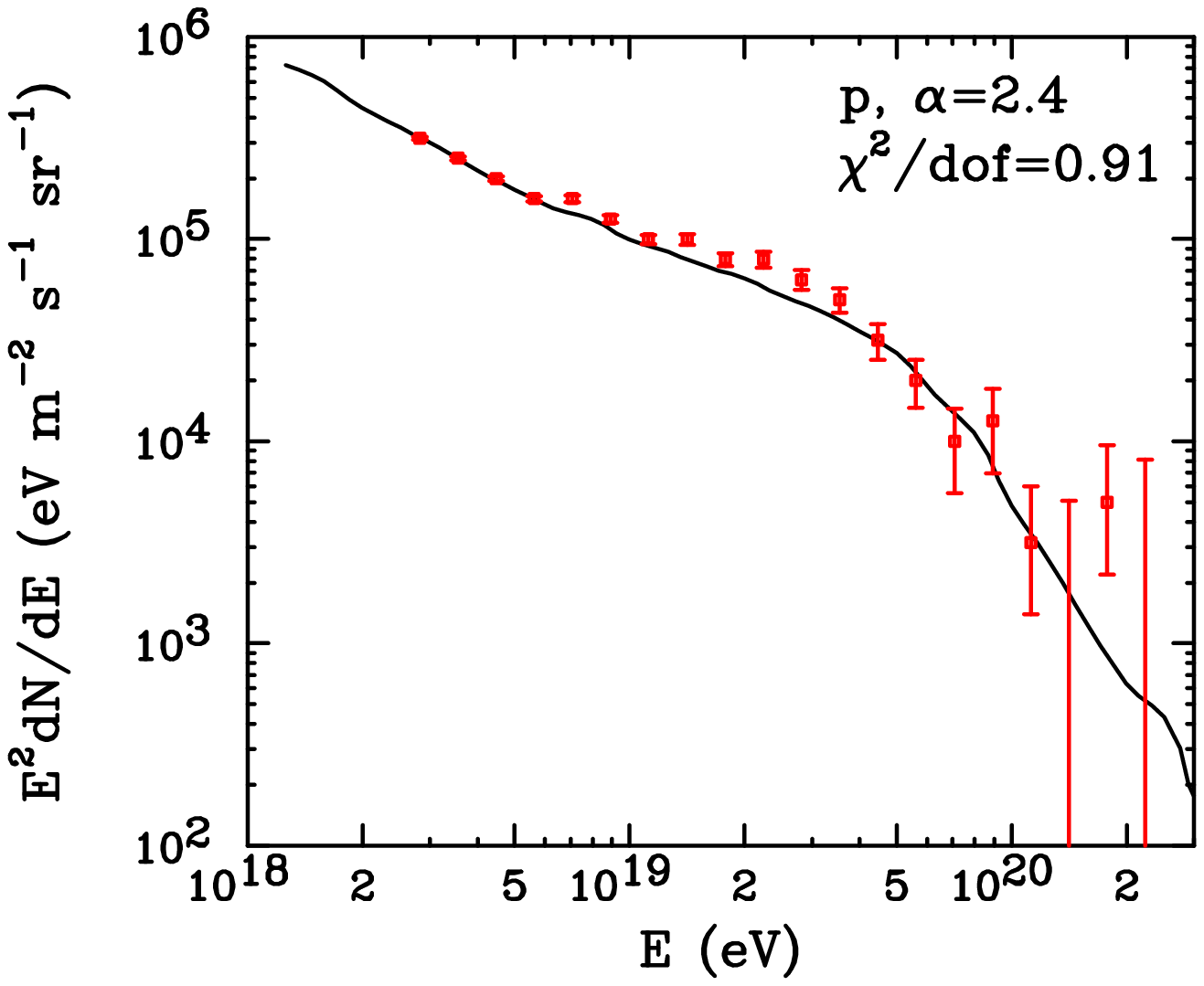,width=.35\columnwidth}
\hspace{1.0cm}
\epsfig{file=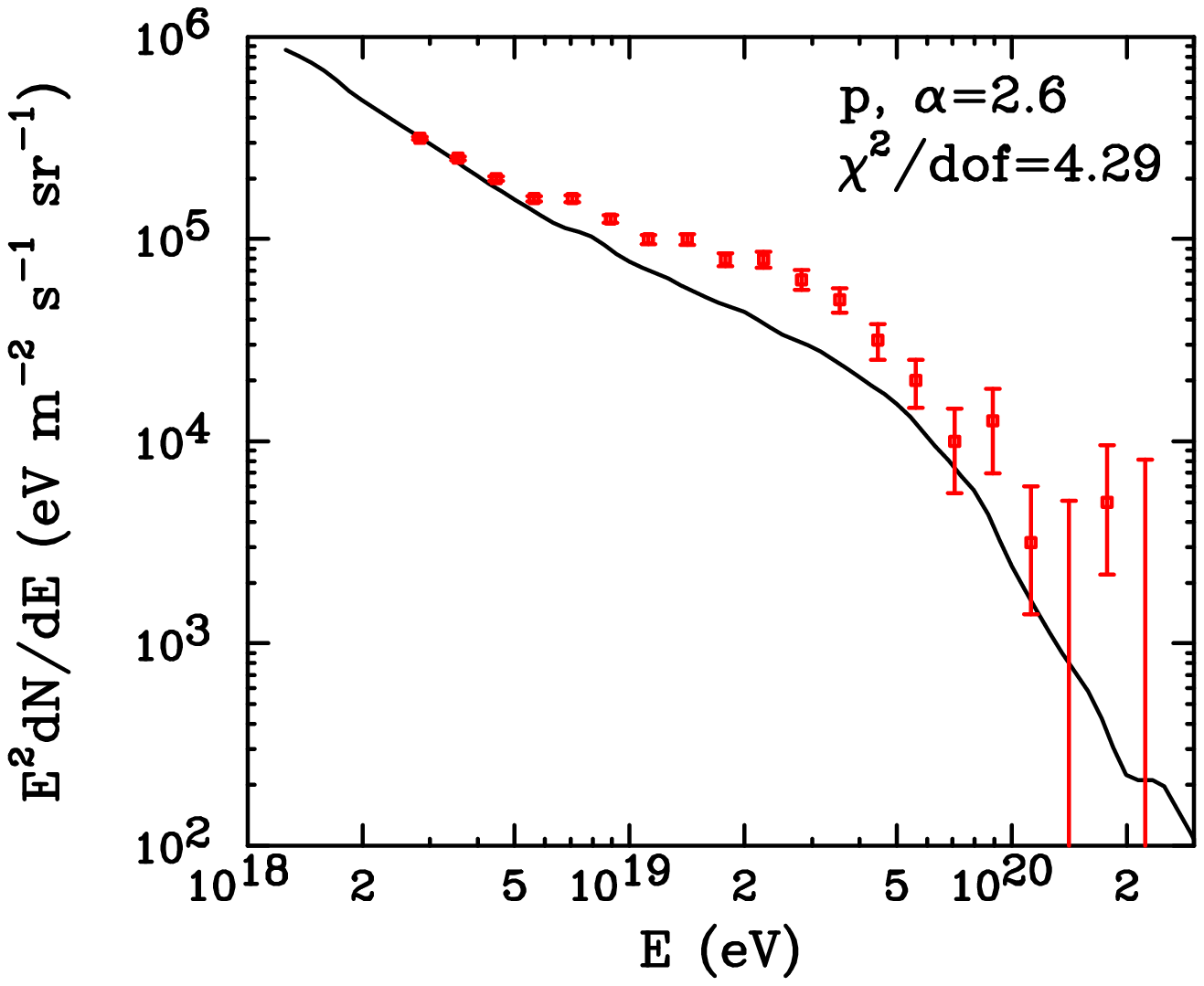,width=.35\columnwidth} \\
\caption{The best fit spectrum for an all-proton UHECR composition for injected spectra with power-law slopes of 2.0, 2.2, 2.4 and 2.6. The dotted, dashed and solid lines denote the
  extragalactic, galactic and combined components, respectively. In
  the lower two frames, the best fit includes a negligible galactic
  component.}
\label{protonspec}
\end{figure}

In this section, we calculate the spectrum and chemical composition of
UHECRs at Earth, for various choices of spectrum and composition as
they are injected at their sources. We then compare our results to the
recent measurements of the PAO to place constraints on the characteristics
of the UHECRs at injection.

First we consider the simple case of an all-proton spectrum. In
Fig.~\ref{protonspec} we show the UHECR spectrum after propagation for
protons injected with a spectrum of $dN/dE \propto E^{-\alpha}$, and
exponentially cutoff above $E_{\rm{max}} = (10^{22} \, \rm{eV})/26$.
(Throughout our study, we adopt an energy cutoff which scales with the
atomic number (electric charge) of the nuclei species, $E_{\rm{max}}
\propto Z$. This is motivated by the fact that, according to the
Hillas criterion~\cite{Hillas:1985is}, cosmic ray sources are able to
accelerate nuclei species to a maximum energy proportional to their
charge. Our definition of $E_{\rm{max}}$ corresponds to the energy
cutoff for iron nuclei, with $Z=26$, and will be accordingly lower for
lighter chemical species.) We have allowed the normalization of the
extragalactic and galactic components of the UHECR spectrum, as well
as the energy at which the galactic component is exponentially cutoff,
to vary freely. In each frame, the spectrum shown is for the choices
of these three parameters which lead to the best statistical fit to
the spectrum measured by the PAO~\cite{Roth:2007in}. Here and throughout
our study, we have adopted a galactic component with a spectral slope
of $\alpha_{\rm{Gal}} = 3.6$ (below the exponential cutoff). We find
that for $\alpha$ in the range of 2.1 to 2.4, an all-proton UHECR
spectrum is consistent with the measured PAO
spectrum~\cite{Roth:2007in} at the 95\% confidence level.

\begin{figure}[!tbp]
\epsfig{file=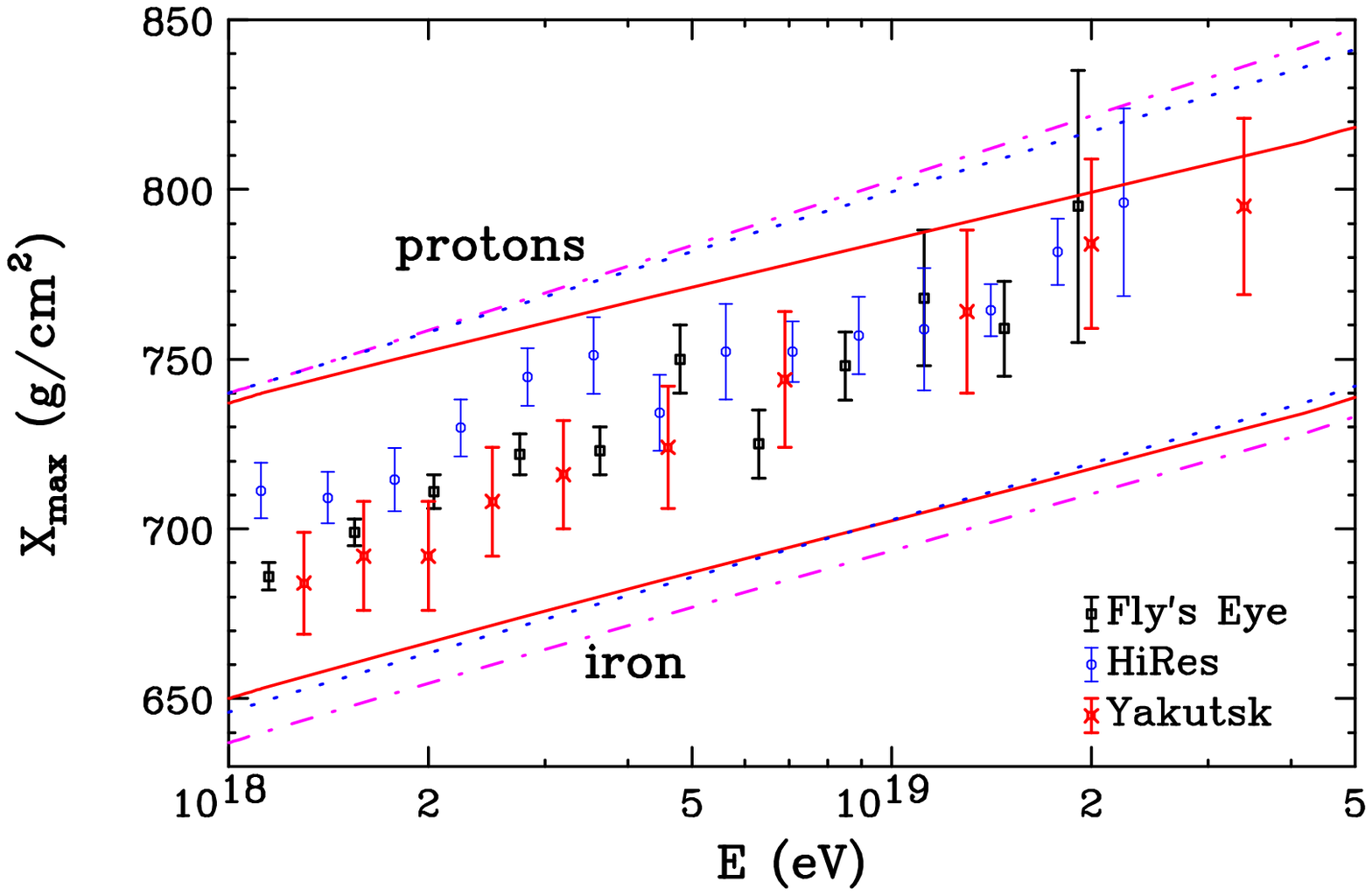,width=.45\columnwidth}
\hspace{0.5cm}
\epsfig{file=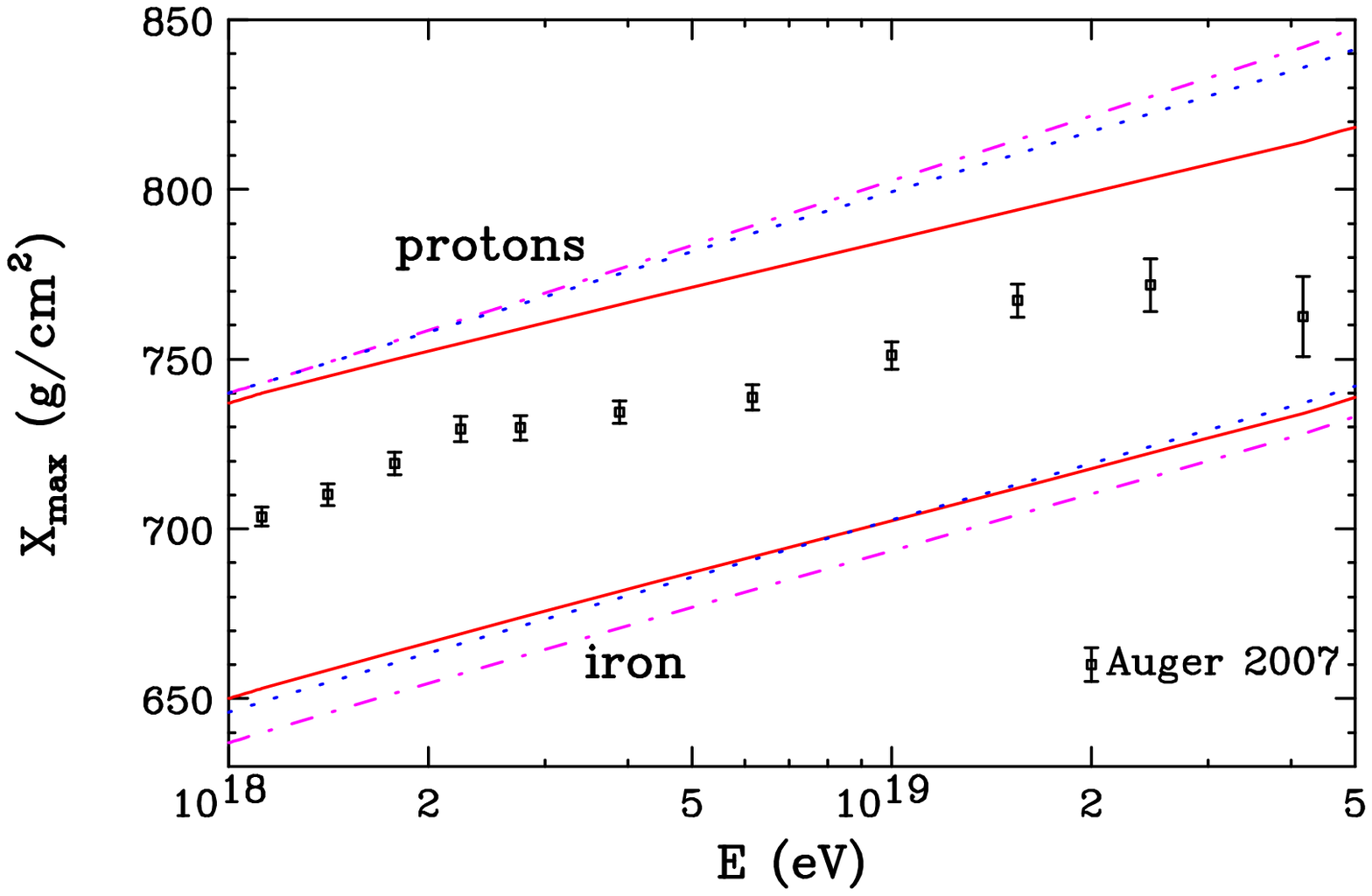,width=.45\columnwidth}
\caption{The values of $X_{\rm{max}}$ as a function of energy as
  measured by Fly's Eye~\cite{Bird:1993yi}, HiRes~\cite{Abbasi:2004nz}
  and Yakutsk~\cite{yakutsk} (left frame) and as measured by the
  PAO~\cite{Unger:2007mc} (right frame). The measurements are compared
  to the predictions for an all-proton and all-iron UHECR composition,
  using three different hadronic physics models. The magenta
  dot-dashed, red solid and blue dotted contours correspond to the
  models EPOS 1.6, QGSJET-III and SIBYLL 2.1, respectively.}
\label{xmax}
\end{figure}

An all-proton spectrum, however, appears to be inconsistent with the
PAO elongation rate measurements~\cite{Unger:2007mc}. In
Fig.~\ref{xmax}, we contrast the PAO measurements of $X_{\rm{max}}$
(right frame) to the measurements of previous UHECR cosmic ray
observatories (left frame), and compare these results to the expected
values for an all-proton or all-iron UHECR composition, as calculated
using three different hadronic physics models (EPOS
1.6~\cite{Werner:2007vd}, QGSJET-III~\cite{Kalmykov:1997te} and SIBYLL
2.1~\cite{Fletcher:1994bd}). Regardless which of these models is used,
the data appears to require the presence of a substantial fraction of
heavy or intermediate mass nuclei in UHECRs. With this in mind, we now
study the shape of the UHECR spectrum as predicted for heavier cosmic
ray species. For details of our treatment of ultrahigh energy cosmic
ray nuclei propagation, see Ref.~\cite{Hooper:2006tn}. For other
studies on this topic, see Ref.~\cite{other}.

\begin{figure}[!tbp]
\epsfig{file=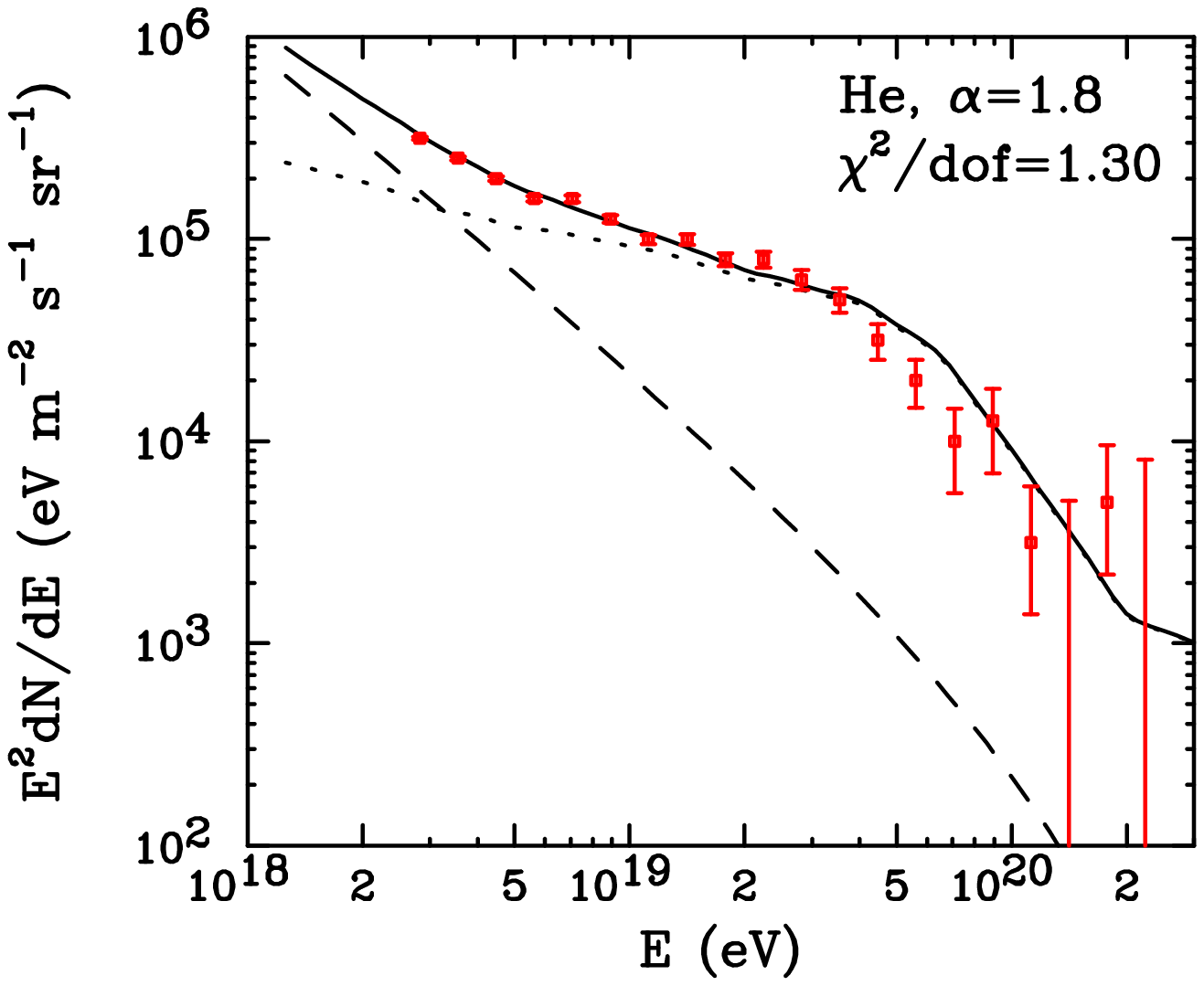,width=.35\columnwidth}
\hspace{1.0cm}
\epsfig{file=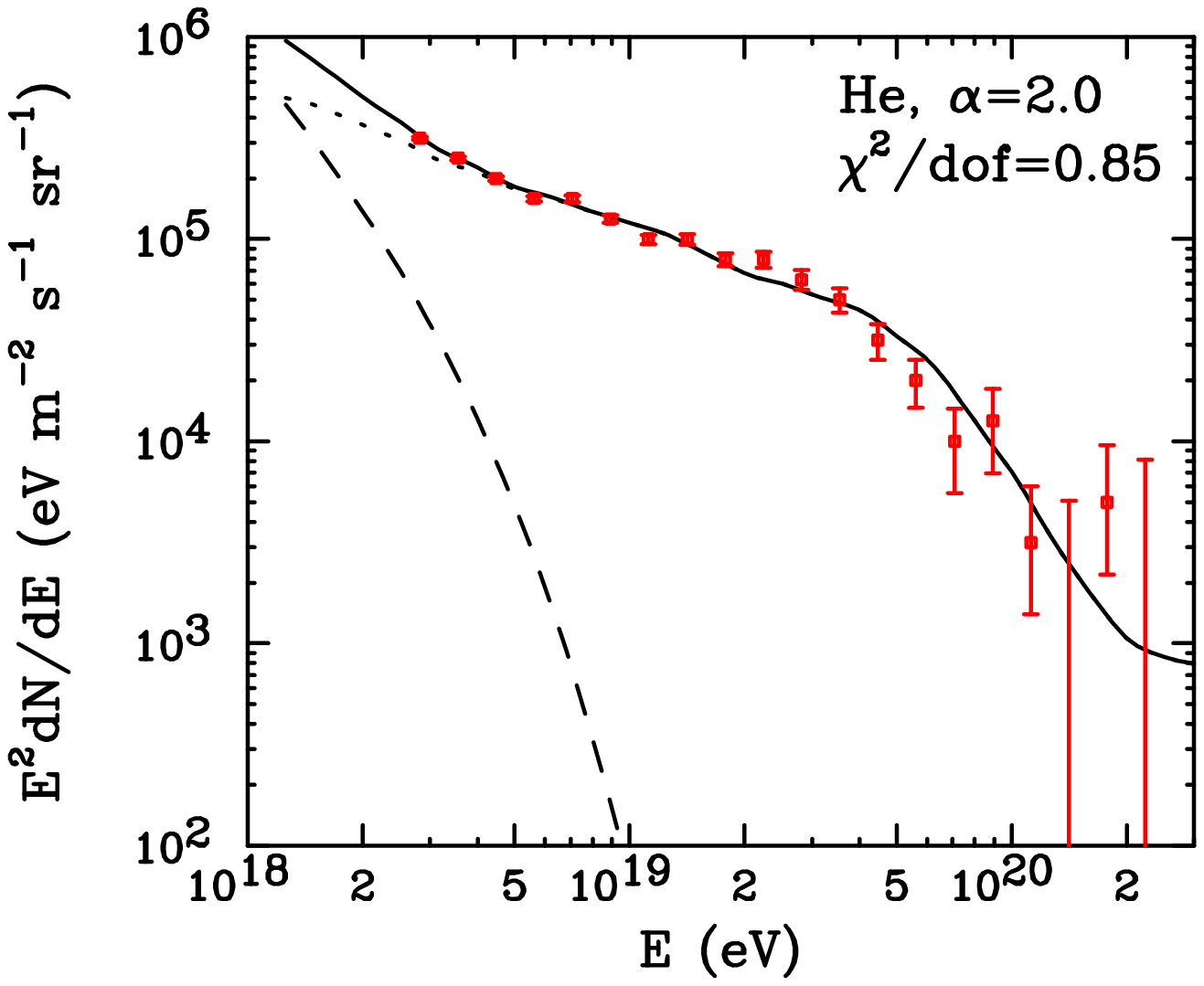,width=.35\columnwidth}\\ 
\epsfig{file=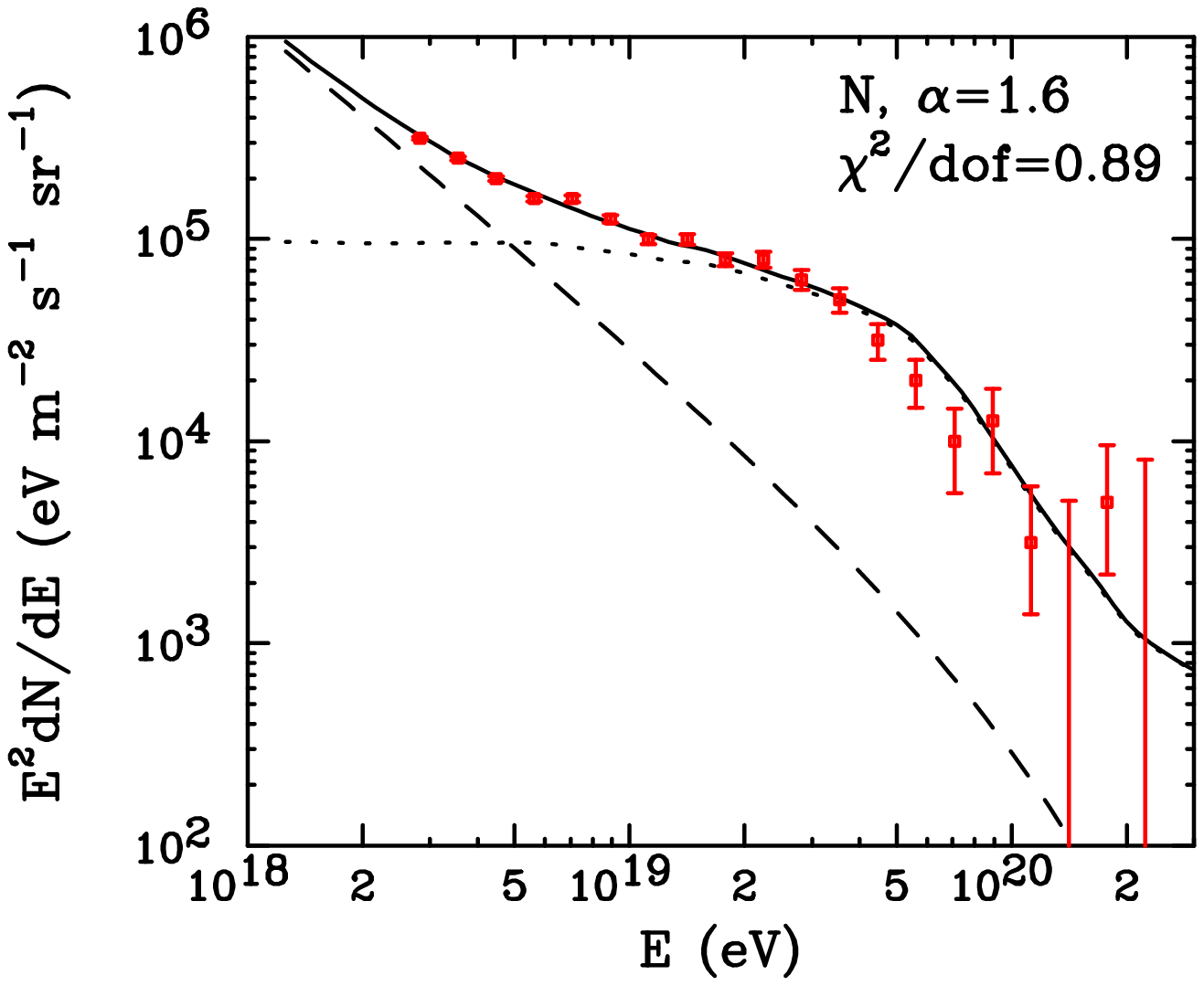,width=.35\columnwidth}
\hspace{1.0cm}
\epsfig{file=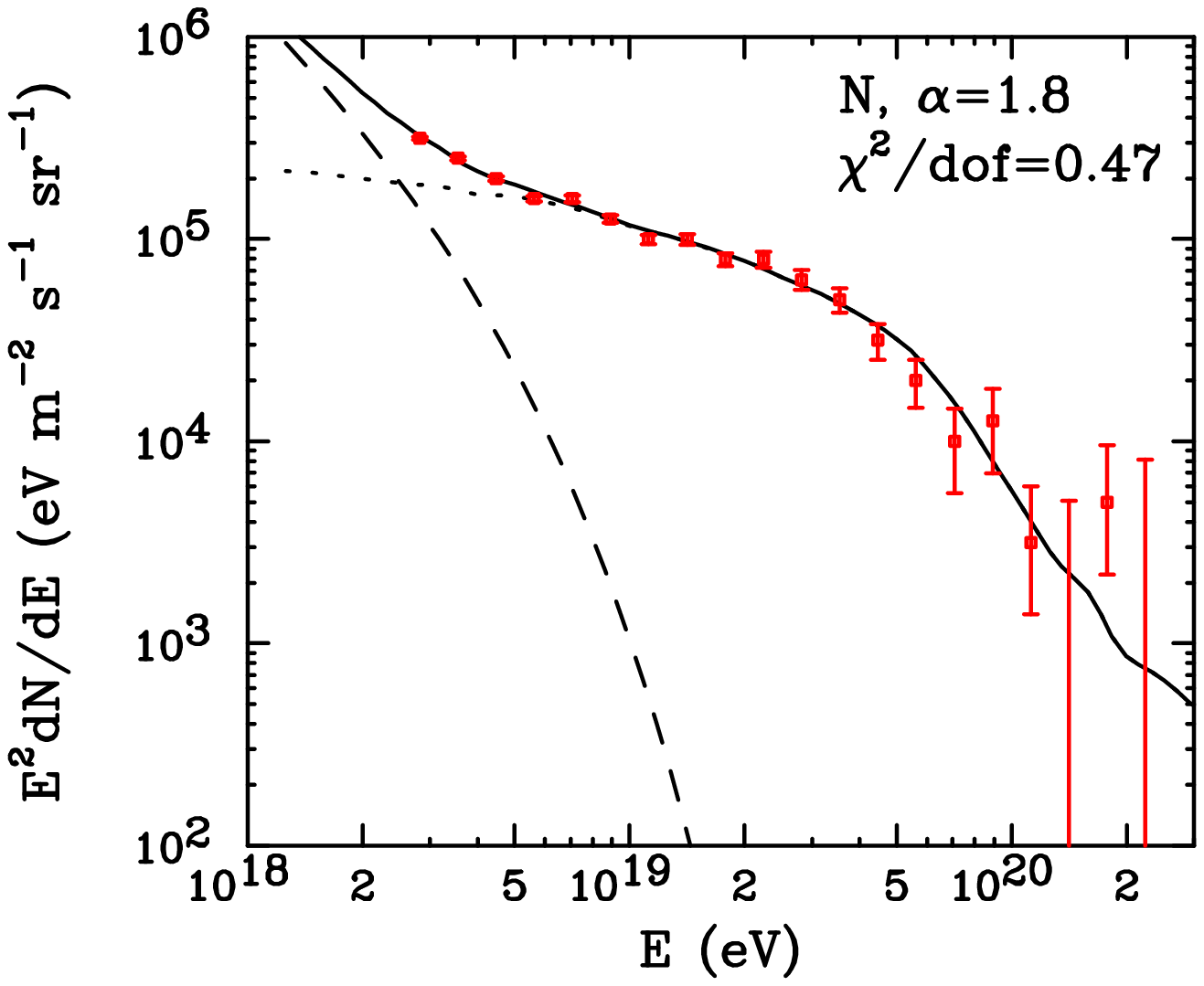,width=.35\columnwidth}\\
\epsfig{file=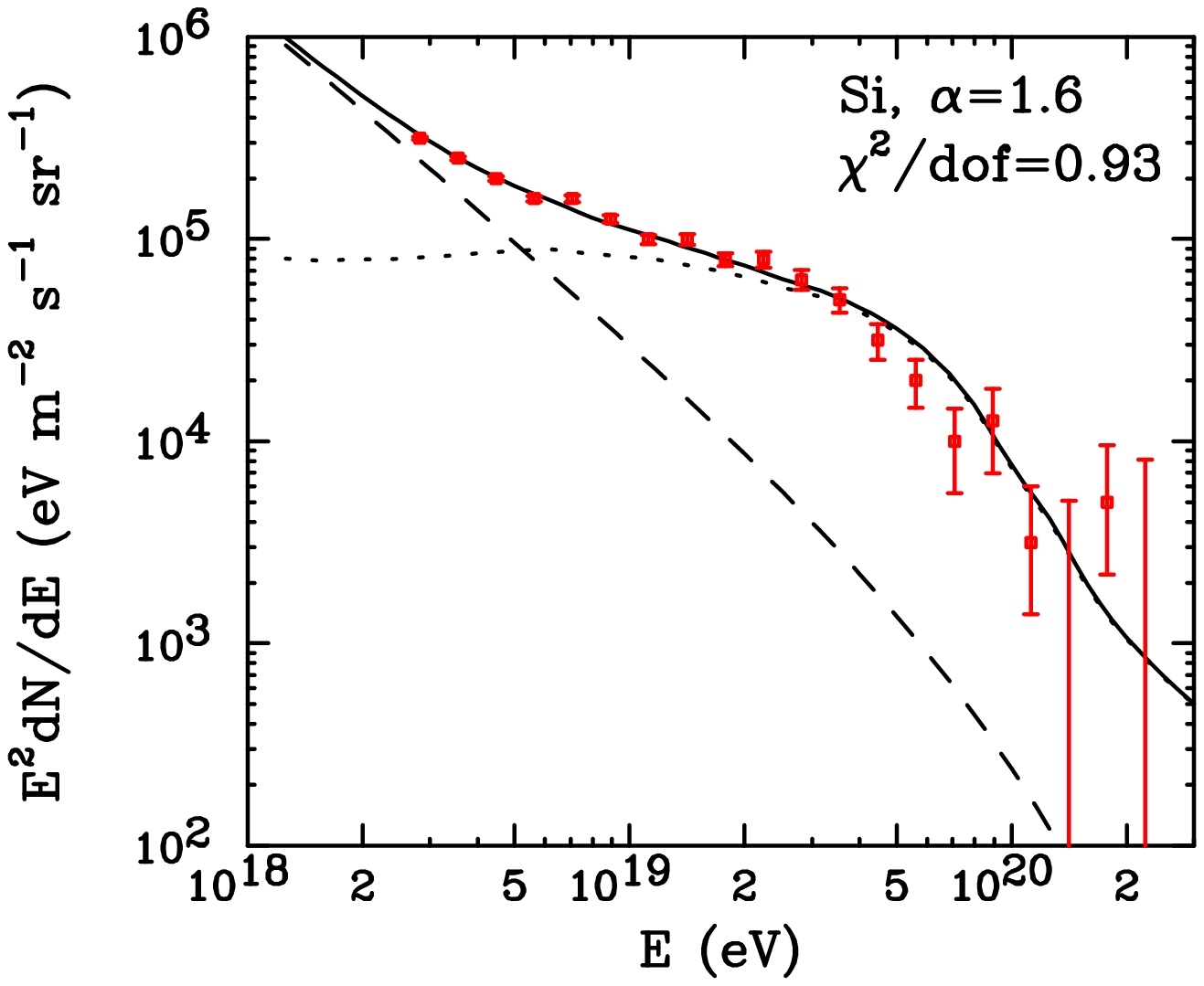,width=.35\columnwidth}
\hspace{1.0cm}
\epsfig{file=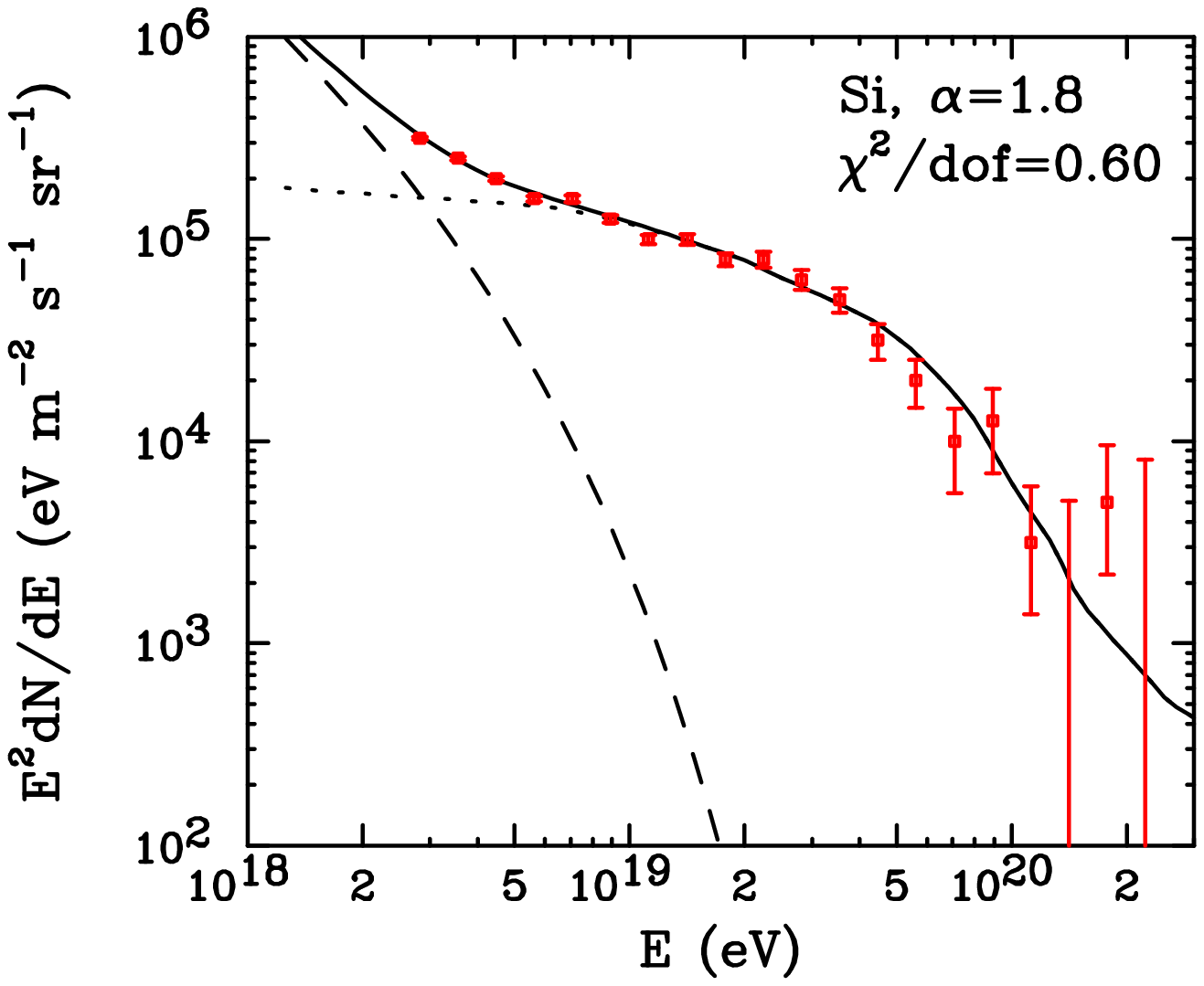,width=.35\columnwidth}\\
\epsfig{file=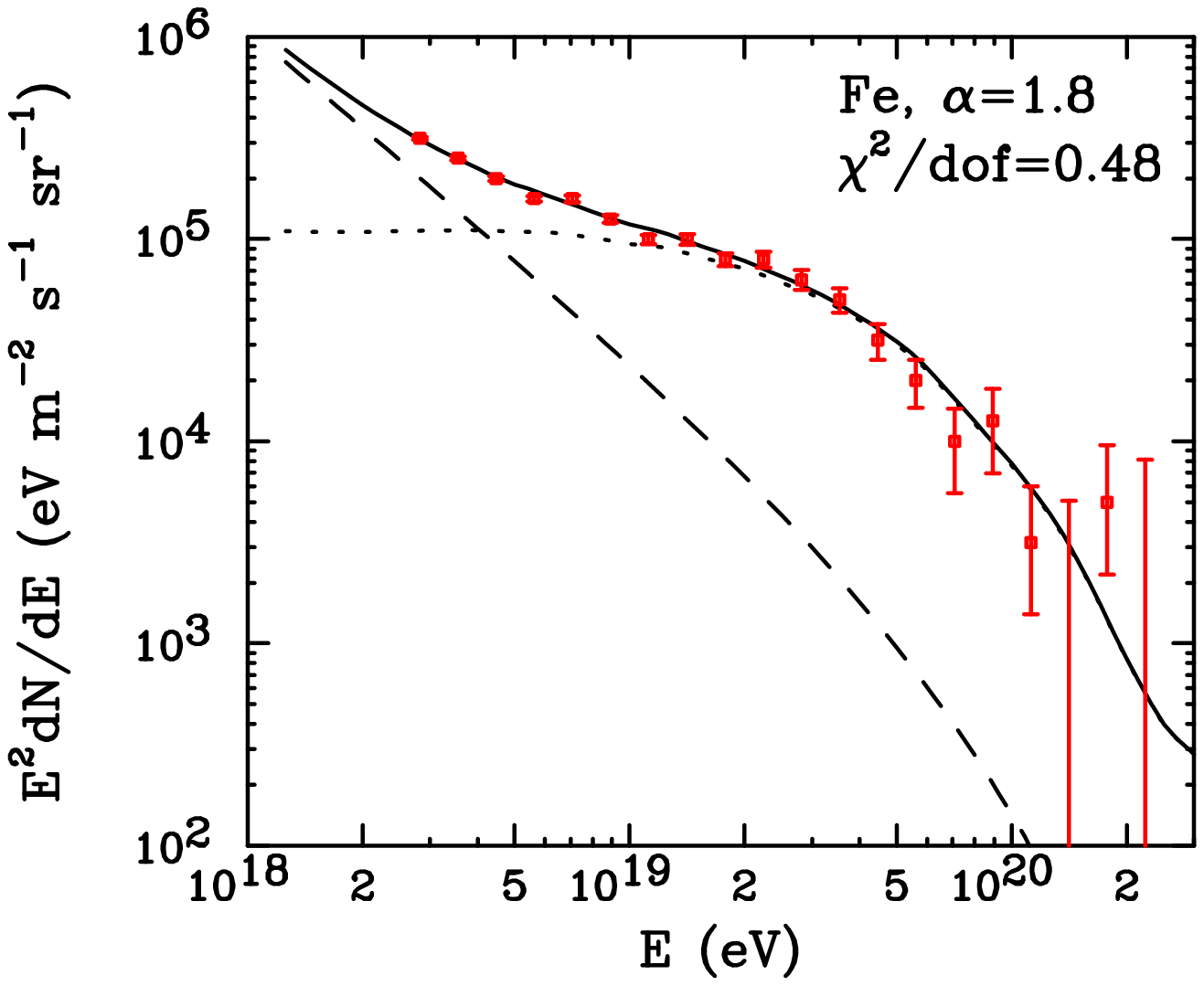,width=.35\columnwidth}
\hspace{1.0cm}
\epsfig{file=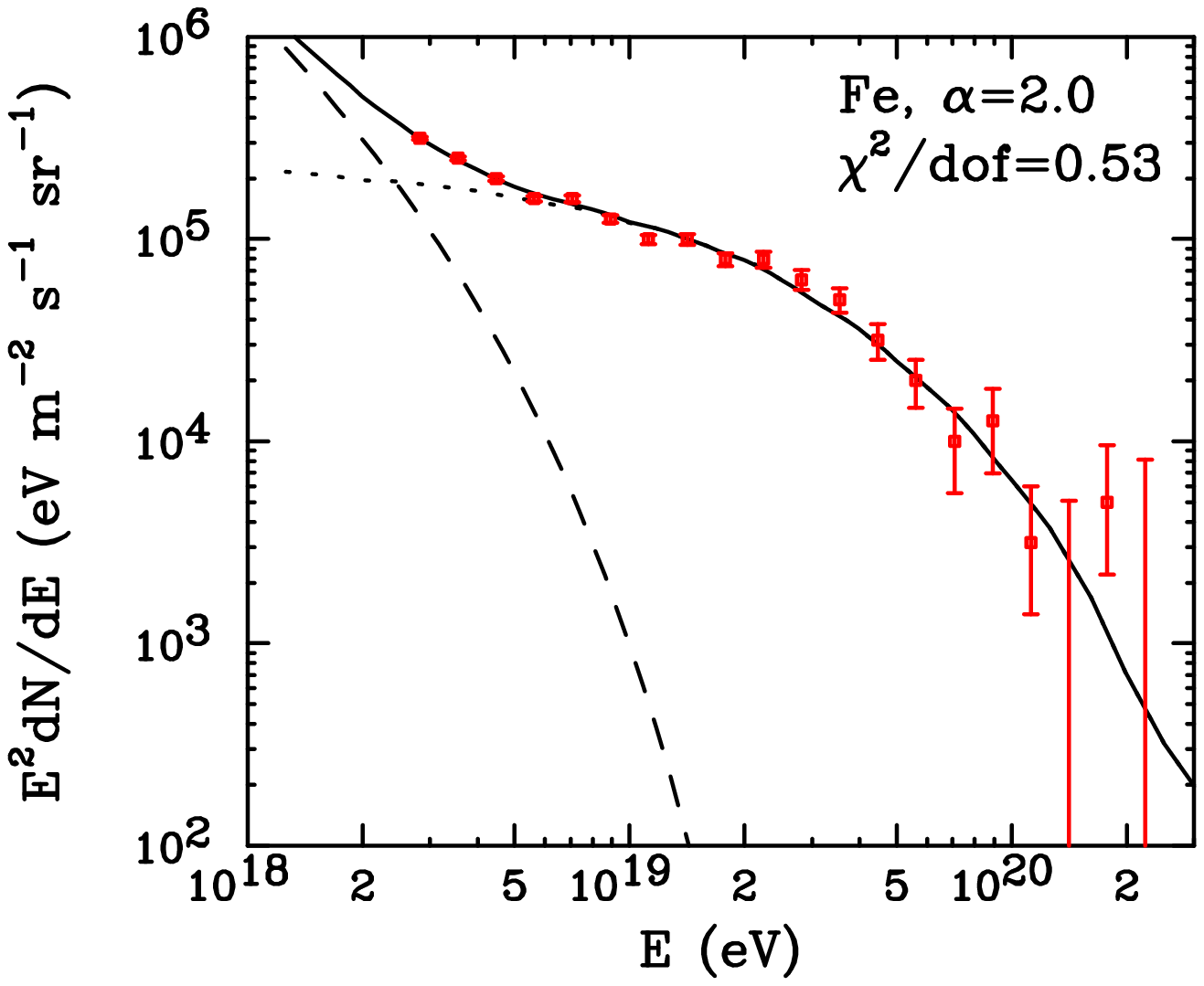,width=.35\columnwidth} 
\caption{The best fit spectrum for an all-helium, all-nitrogen,
  all-silicon or all-iron UHECR composition (at injection) for
  different injected power-law slopes. The dotted, dashed and solid
  lines denote the extragalactic, galactic and combined components,
  respectively.} \label{nucleispec}
\end{figure}

In Fig.~\ref{nucleispec}, we show the UHECR spectrum after propagation
for an injected composition consisting entirely of helium, nitrogen,
silicon or iron. We show results for various values of the spectral
index, $\alpha$, and have taken $E_{\rm{max}} = (10^{22} \,
\rm{eV})\times (Z/26)$, where $Z$ is the electric charge of the
injected species of nuclei. In each case, we find that smaller values
of $\alpha$ are preferred than for an all-proton spectrum. In particular, a spectral slope
of 1.6--2.1 is able to provide good fits, as opposed to 2.1--2.4 for
protons.

\begin{figure}[!tbp]
\epsfig{file=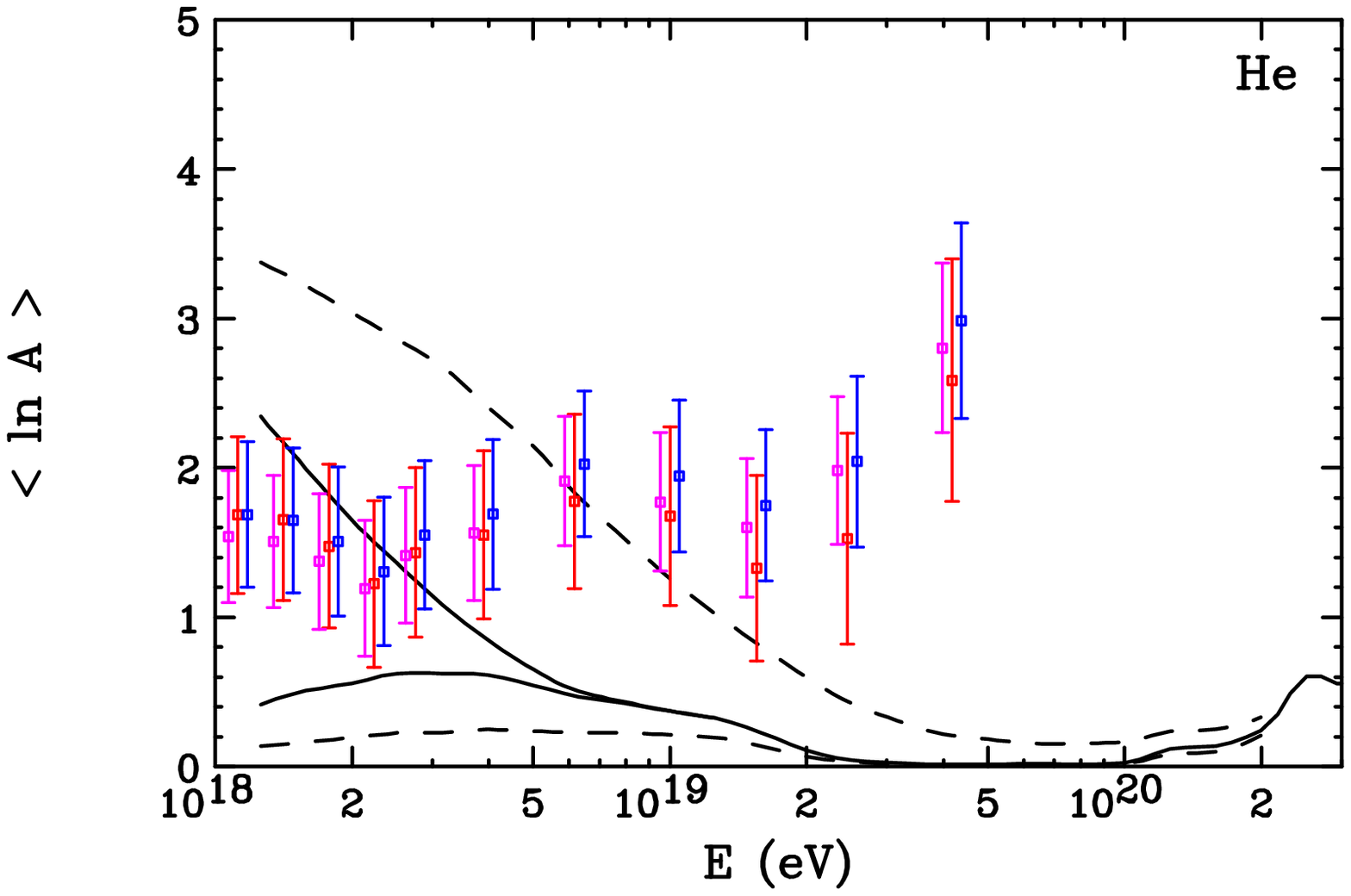,width=.35\columnwidth}
\hspace{1.0cm}
\epsfig{file=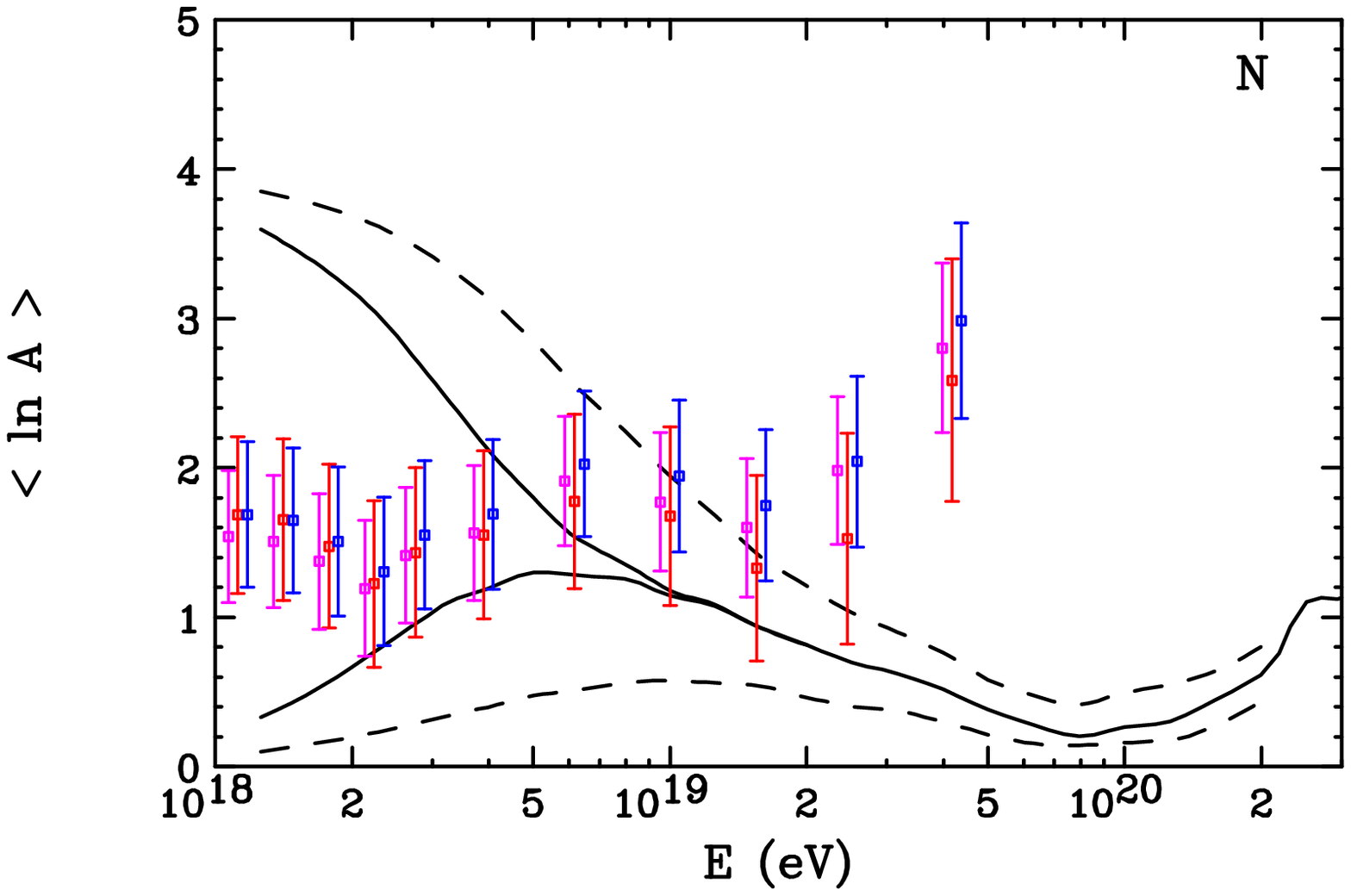,width=.35\columnwidth}\\
\vspace{0.5cm}
\epsfig{file=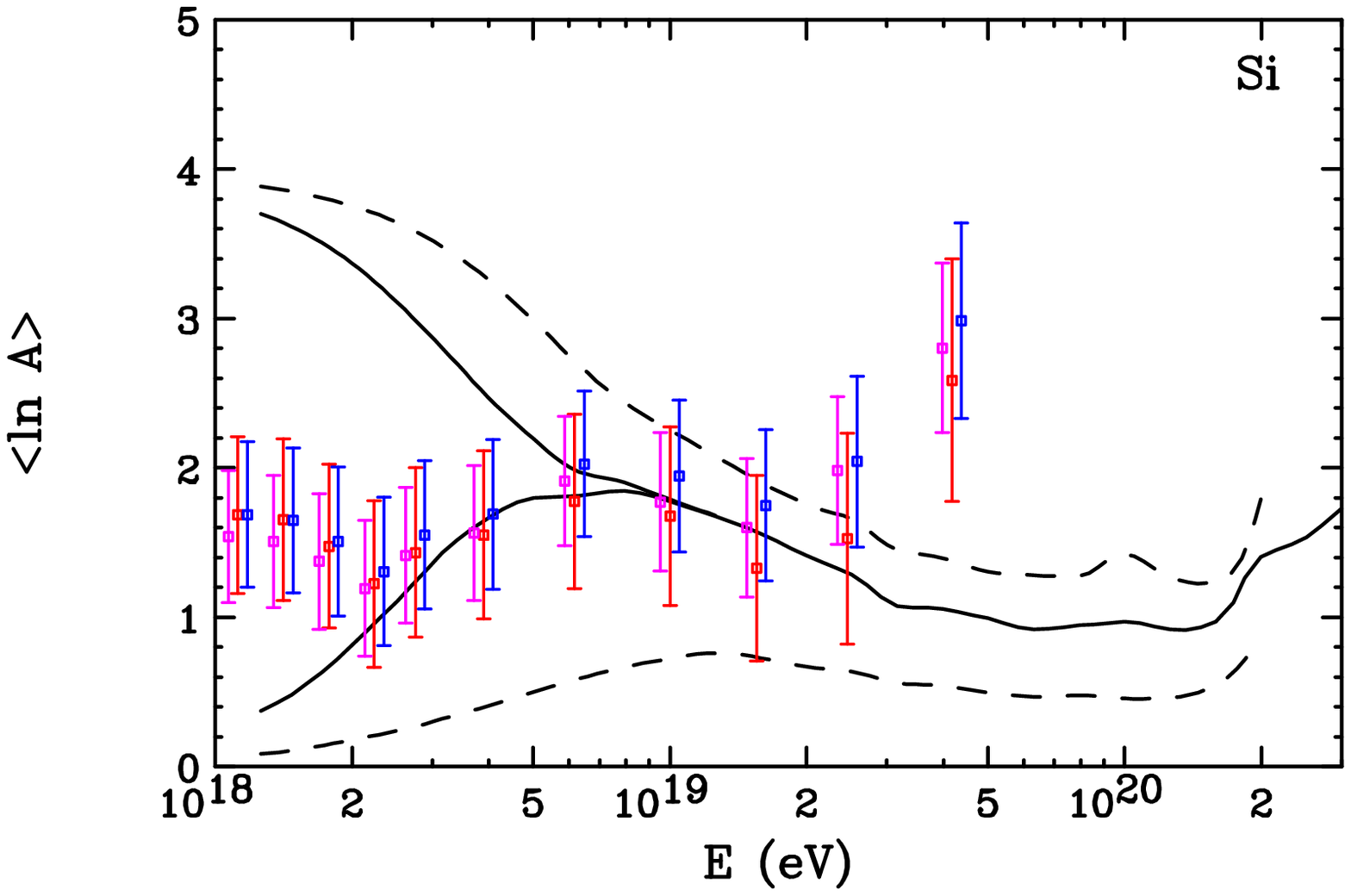,width=.35\columnwidth}
\hspace{1.0cm}
\epsfig{file=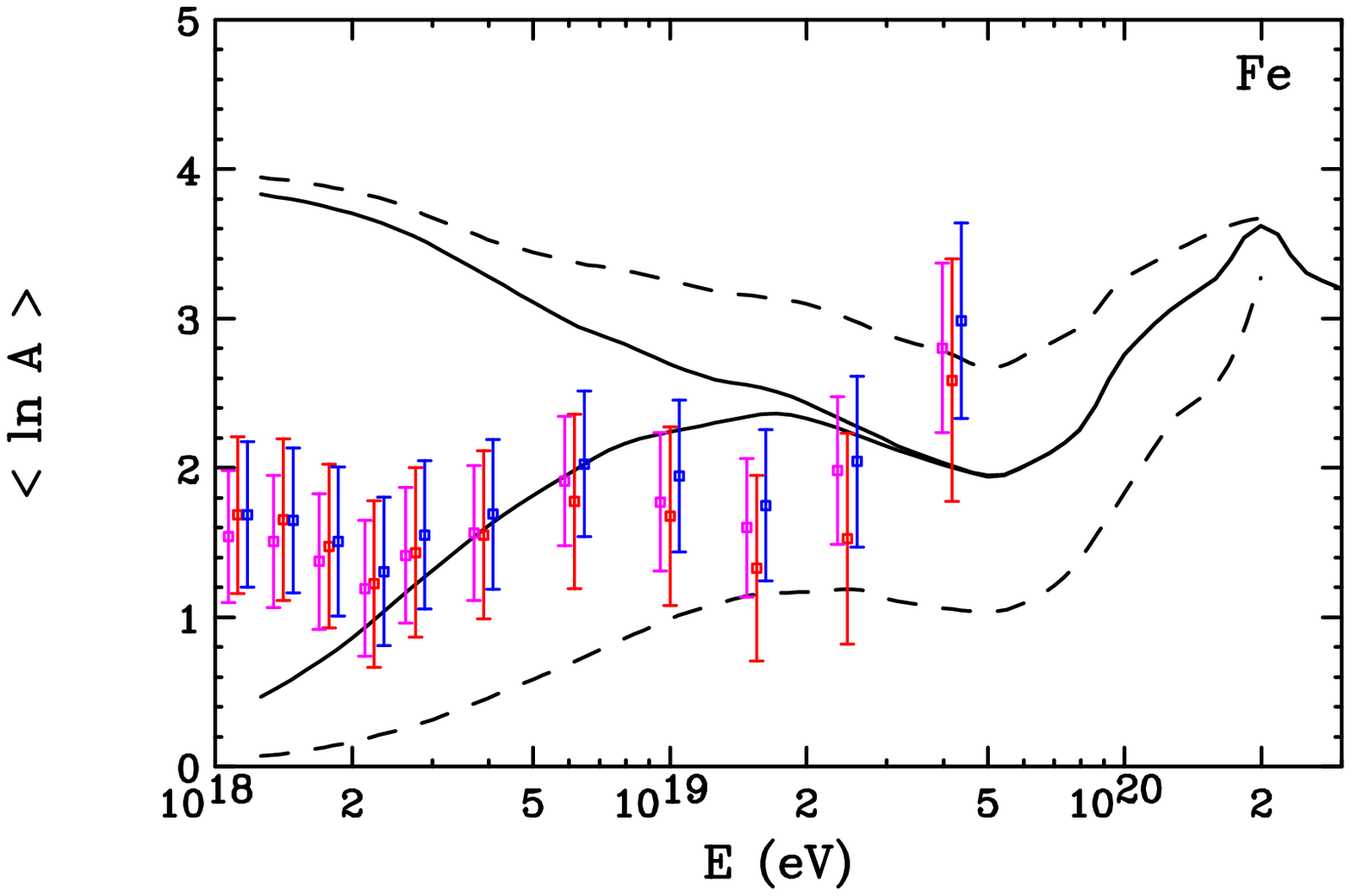,width=.35\columnwidth}
\caption{The average composition at Earth for pure helium, nitrogen,
  silicon or iron nuclei (at injection). The solid lines denote the model
best fit to the PAO spectrum, with the range shown below $\sim 10^{19}$ eV
  resulting from variations in the composition of the galactic
  component. The range denoted by the dashed lines is that which is
  consistent with the spectrum measured by the PAO (within the 95\%
  confidence level). In each case, an exponential cutoff above
  $E_{\rm{max}} = 10^{22}\, \rm{eV} \times (Z/26)$ was assumed. The
  error bars correspond to measurements of the elongation rate by the PAO
  using EPOS 1.6 (magenta, offset left), QGSJET-III (red, center) and
  SIBYLL 2.1 (blue, offset right), and in each case include both
  statistical and systematic uncertainties.}
\label{nucleicomp}
\end{figure}

We have also calculated the average composition of the UHECR spectrum
for each of these nuclear species, after the effects of propagation are
accounted for. In Fig.~\ref{nucleicomp}, we have plotted the average
of the logarithm of the atomic mass of the UHECRs, as a function of
energy, for various choices of injected nuclei species. The solid
lines denote the result using the parameters (spectral index and
energy cutoff of the extragalactic component and energy cutoff of the
galactic contribution) which provide the best fit to the PAO's spectrum
measurement~\cite{Roth:2007in}. The dashed lines denote the range over
which we find good fits to the PAO spectrum (within the 95\%
confidence level). We have made no assumptions regarding the galactic
composition, which results in a wide range of possible average
compositions below $\sim 10^{19}$~eV.

In Fig.~\ref{nucleicomp}, the error bars shown correspond to the PAO
measurements of the elongation rate~\cite{Unger:2007mc}.  To translate
between this quantity and the average of the logarithm of the
composition ($<\ln A>$), a model of the hadronic physics involved must
be adopted. In the figure, we show results corresponding to three
hadronic models (EPOS 1.6~\cite{Werner:2007vd},
QGSJET-III~\cite{Kalmykov:1997te} and SIBYLL
2.1~\cite{Fletcher:1994bd}). The error bars shown denote both the
statistical errors and systematic uncertainties.

It should be noted that the interactions of the highest energy cosmic
rays occur with center-of-mass energies well above those probed at
collider experiments. These models are thus based on
extrapolations from lower energy accelerator data. In the future, it
will become increasingly possible to distinguish between the effects
of chemical composition and currently unknown hadronic physics through
the detailed analysis of all observables related to composition (e.g.,
elongation rate, spread of $X_{\rm max}$) together with energy
spectrum and anisotropy measurements. Because of the limited statistics
available at these extreme energies, a definitive analysis including anisotropy
measurements~\cite{Anchordoqui:2007tb} will have to await several years
of data accumulation.

So far, we have fixed the exponential cutoff in the injected spectrum
of UHECRs to $E_{\rm{max}} = 10^{22}\, \rm{eV} \times (Z/26)$, where
$Z$ is the atomic number of the nuclei species. If the value of
$E_{\rm{max}}$ is lowered, we find that the injected spectrum must be
further hardened to fit the spectrum measured by the PAO. For
$E_{\rm{max}} = 10^{21}\, \rm{eV} \times (Z/26$), for example, we find
that protons and iron require $\alpha$ in the range of 1.4--1.9 and
1.4--2.0, respectively. More generally, lower values of $E_{\rm max}$
require harder spectral indices. We did not consider values of
$\alpha$ below 1.4. We found no acceptable fits for helium, nitrogen
or silicon with this choice of $E_{\rm{max}}$. If we consider
$E_{\rm{max}} = 10^{20}\, \rm{eV} \times Z/26$, we find an acceptable
fit only for iron, with $\alpha$=1.4--1.7.

\begin{figure}[!tbp]
\epsfig{file=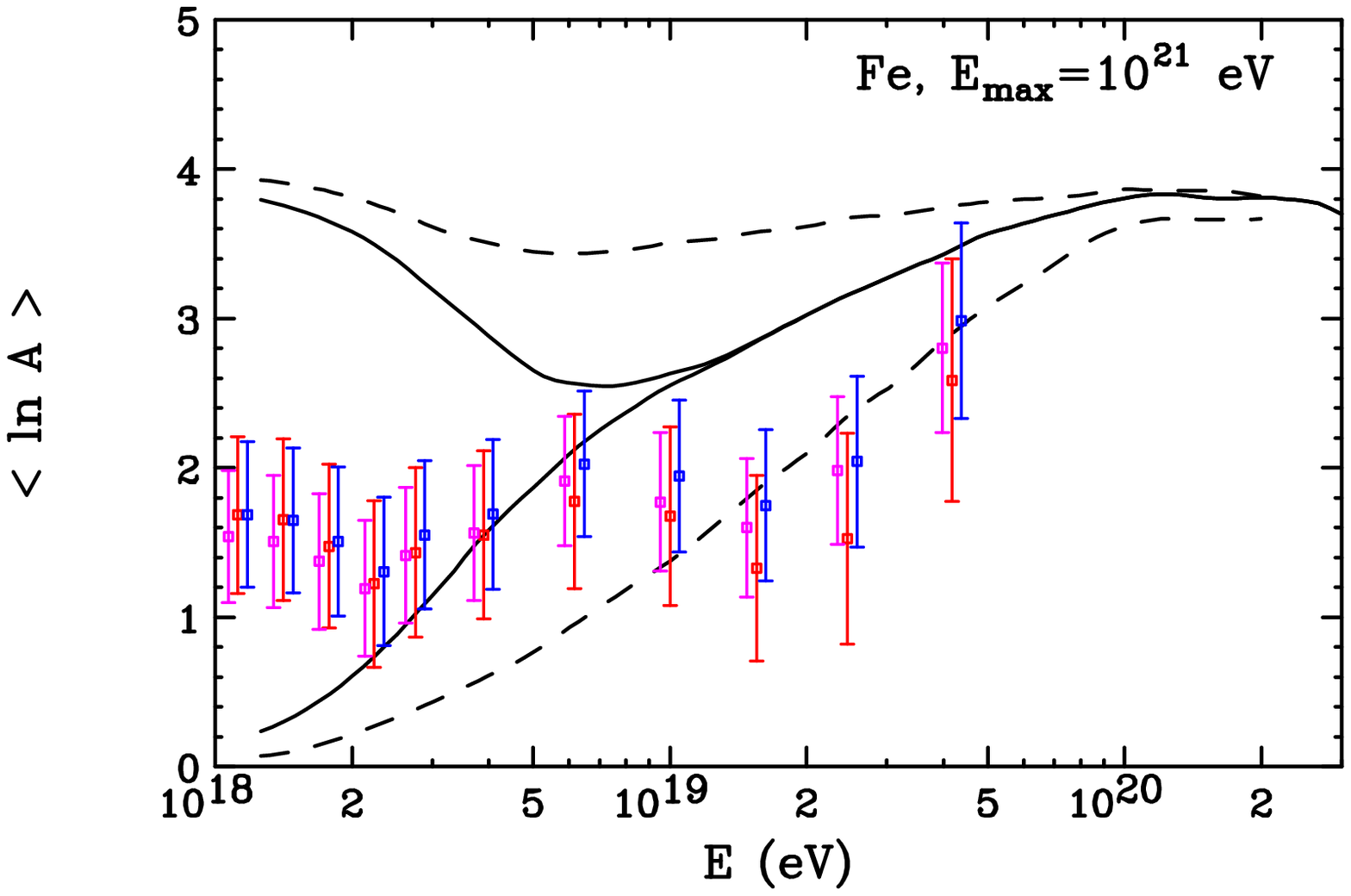,width=.35\columnwidth}
\hspace{0.5cm}
\epsfig{file=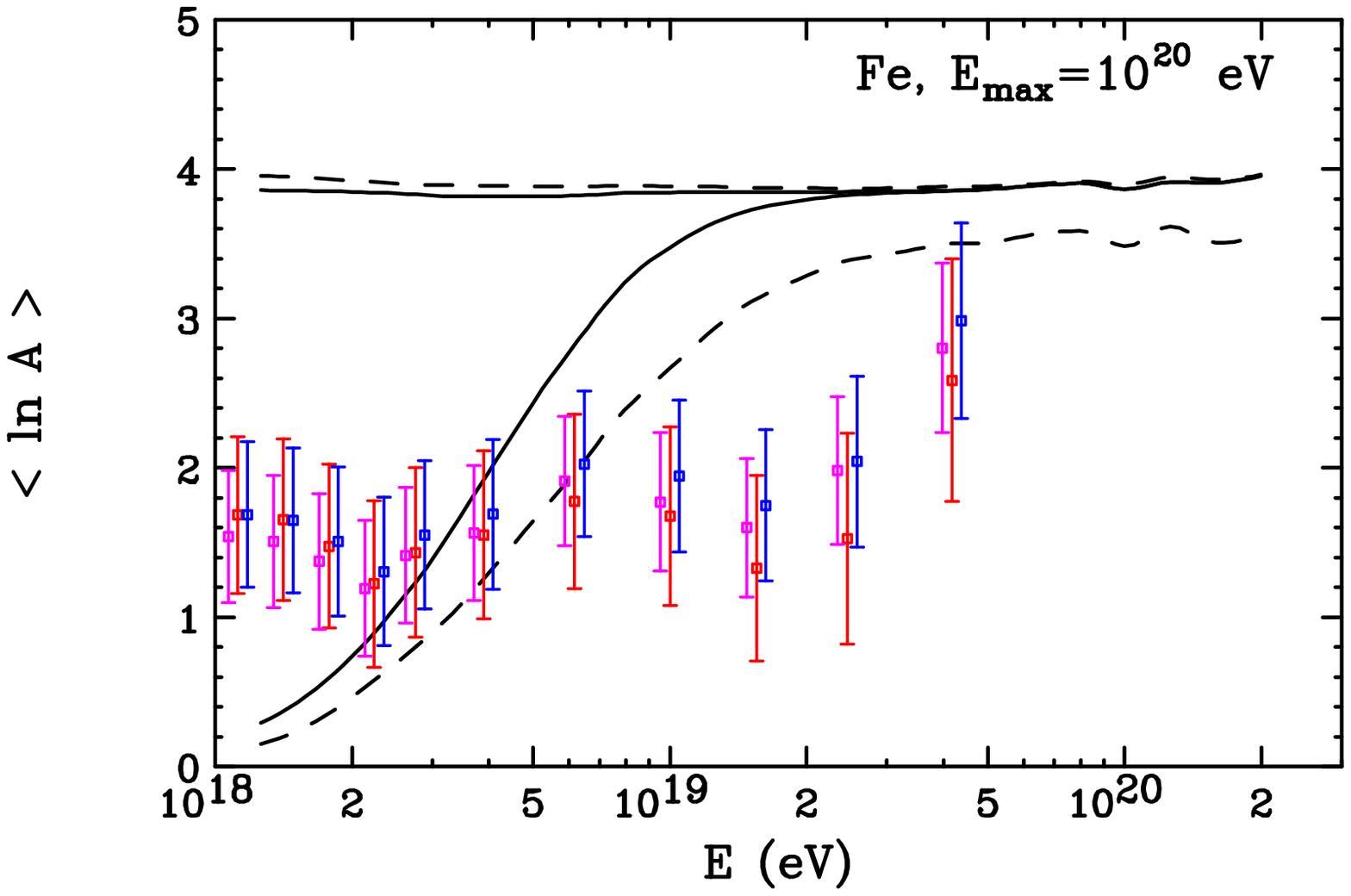,width=.35\columnwidth}
\caption{The average composition at Earth for a pure iron UHECR
  spectrum at injection, for two values of the exponential cutoff in
  energy. The solid lines denote the best fit model, with the split
  below $\sim 10^{19}$ eV resulting from variations in the composition
  of the galactic component. The range denoted by the dashed lines is
  that which is consistent with the spectrum measured by the PAO (within
  the 95\% confidence level). The error bars denote the $X_{\rm{max}}$
  measurements using EPOS 1.6 (magenta, offset left), QGSJET-III (red,
  center) and SIBYLL 2.1 (blue, offset right).}
\label{emaxcomp}
\end{figure}

The value of the energy cutoff adopted also affects the average UHECR
composition which is expected to be observed at Earth. In
Fig.~\ref{emaxcomp}, we show the expected composition for an all-iron
injected spectrum with $E_{\rm{max}} = 10^{21}\, \rm{eV}$ and
$E_{\rm{max}} = 10^{20}\, \rm{eV}$. Lowering the maximum energy to
which UHECRs are accelerated results in a considerably heavier
composition at Earth.

\begin{figure}[!tbp]
\epsfig{file=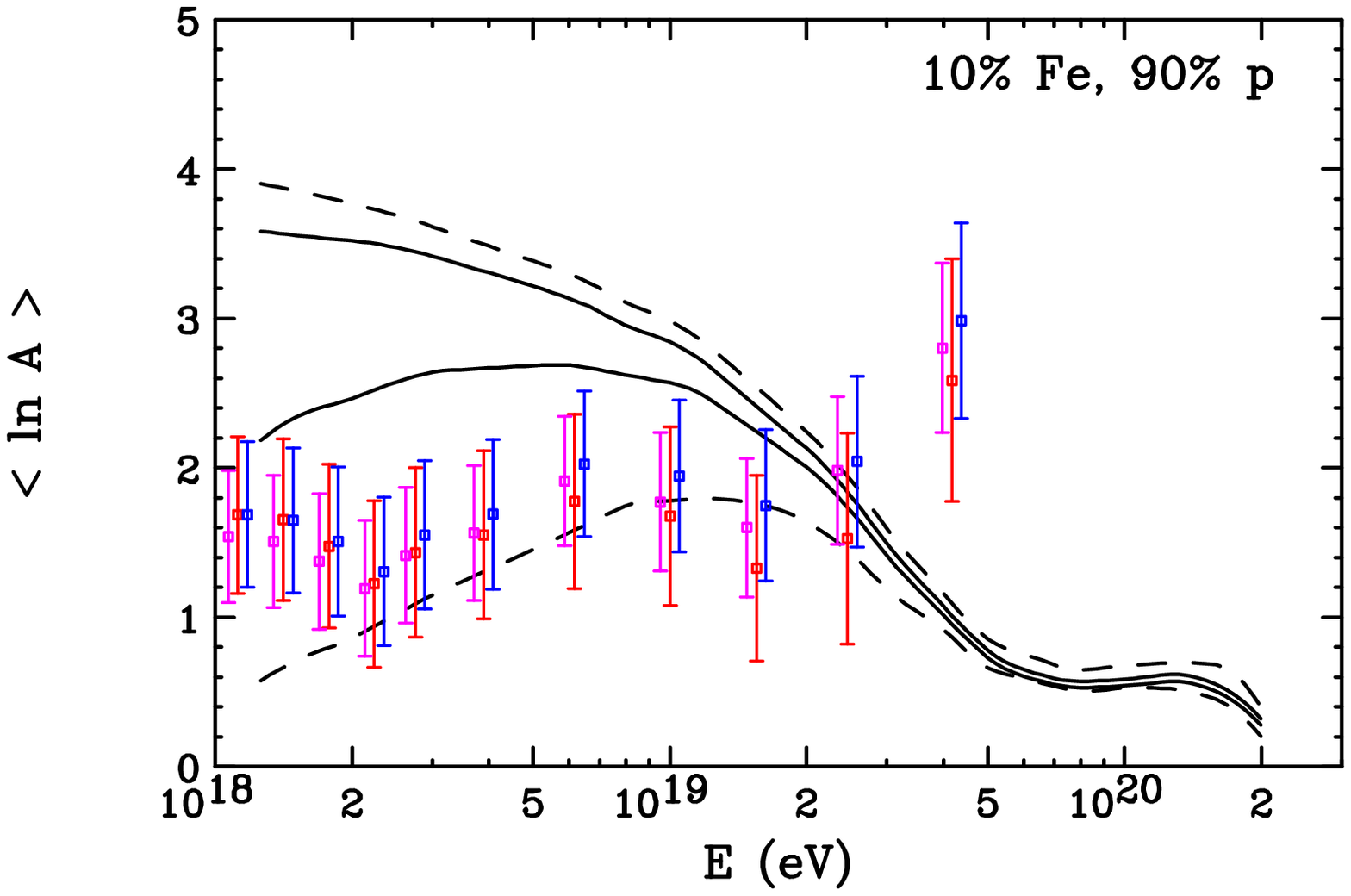,width=.35\columnwidth}
\hspace{0.5cm}
\epsfig{file=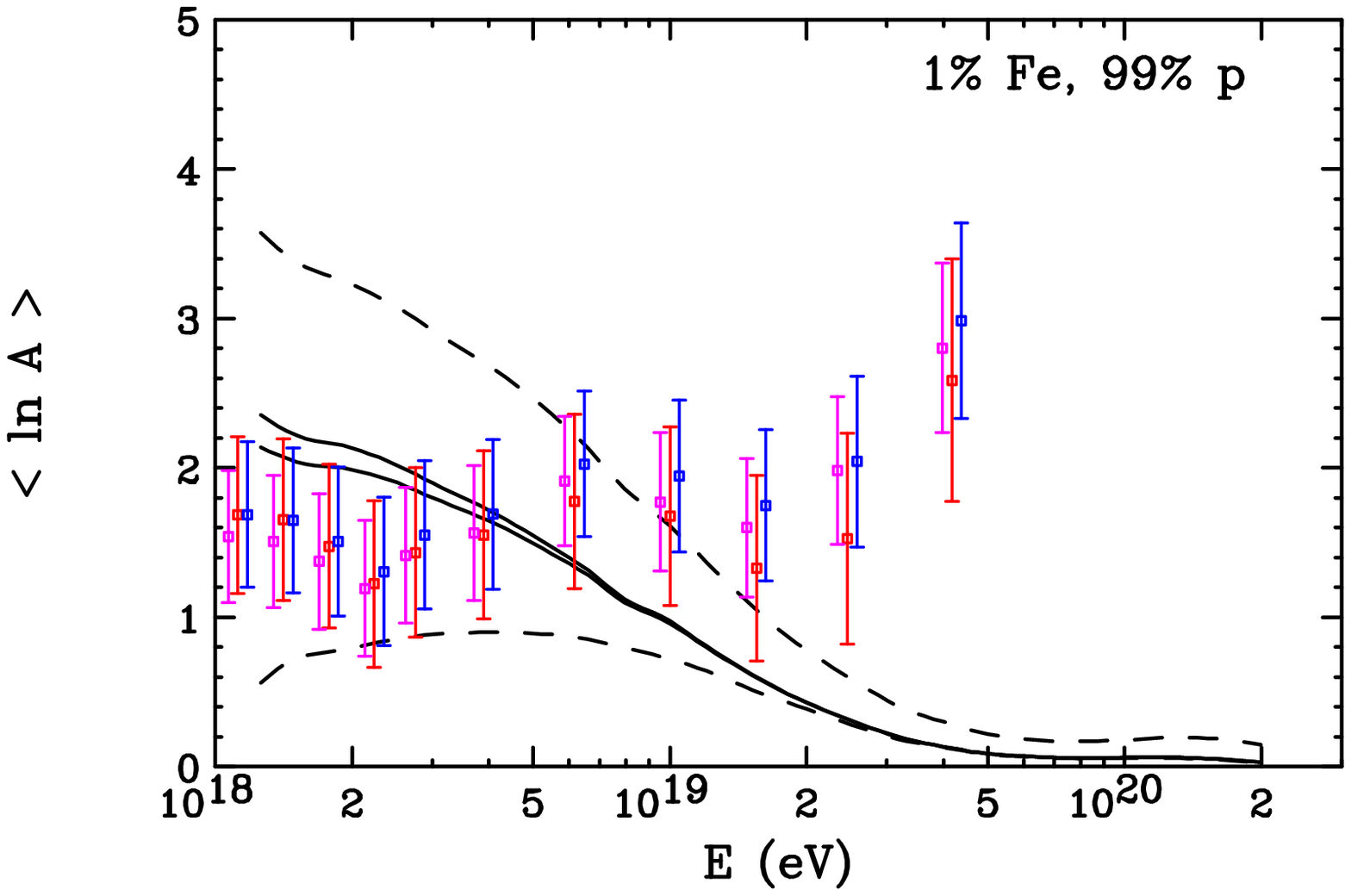,width=.35\columnwidth}
\caption{The average composition at Earth for two mixtures of iron
  nuclei and protons in the injected spectrum. In each frame, an
  exponential energy cutoff of $E_{\rm{max}} = 10^{22}\, \rm{eV}
  \times (Z/26$) has been used. The solid lines denote the best fit
  model, with the split below $\sim 10^{19}$ eV resulting from
  variations in the composition of the galactic component. The range
  denoted by the dashed lines is that which is consistent with the
  spectrum measured by the PAO (within the 95\% confidence level). The
  error bars denote the $X_{\rm{max}}$ measurements using EPOS 1.6
  (magenta, offset left), QGSJET-III (red, center) and SIBYLL 2.1
  (blue, offset right).} \label{mix}
\end{figure}

Of course, the UHECR spectrum can also be generated through the
injection of a mixture of protons and various species of nuclei. In
Fig.~\ref{mix}, we plot the average composition at Earth found for
mixtures of iron nuclei and protons injected in sources of UHECRs. We
find that even a very small fraction of injected iron nuclei can
dominate the observed spectrum and composition at Earth, essentially
because ultrahigh energy heavy nuclei, on average, travel further than
protons.

After considering both the spectrum~\cite{Roth:2007in} and elongation
rate~\cite{Unger:2007mc} measurements made by the PAO, we find a
fairly wide range of spectral indices and compositions which are
consistent with the observations. In considering the composition, we
consider only the 5 highest energy bins, where the extragalactic
component of the UHECR spectrum is most likely to dominate. Varying
the injected spectral index ($\alpha$), the exponential energy cutoff
($E_{\rm{max}}$) and the injected chemical composition, we find the
following results. An all-iron injected spectrum can fit the data well
for $\alpha=$1.4-2.1, depending on the choice of $E_{\rm{max}}$ which
is adopted. All-silicon and all-nitrogen spectra can fit for
$\alpha=$1.6-2.0 and 1.6-1.9, respectively. An all-helium or
all-proton injection spectrum is, however, not consistent with the
data.

Protons with an admixture of as little as $3\%$ ($1\%$) iron or $7\%$
($2\%$) silicon, can be consistent with all measurements for
$E_{\rm{max}} = 10^{22}\, \rm{eV}\, (10^{21} \, \rm{eV}) \times Z/26$.
Nitrogen is consistent only if injected in equal or greater amount as
protons.


\section{Predictions For The Cosmogenic Neutrino Spectrum}

Ultrahigh energy protons above the ``GZK cutoff'' \cite{gzk} interact
with the cosmic microwave and infrared backgrounds as they propagate
over cosmological distances.  These interactions generate pions and
neutrons, which decay to produce neutrinos. The accumulation of these
neutrinos over cosmological time is known as the cosmogenic neutrino
flux.

\begin{figure}[!tbp]
\epsfig{file=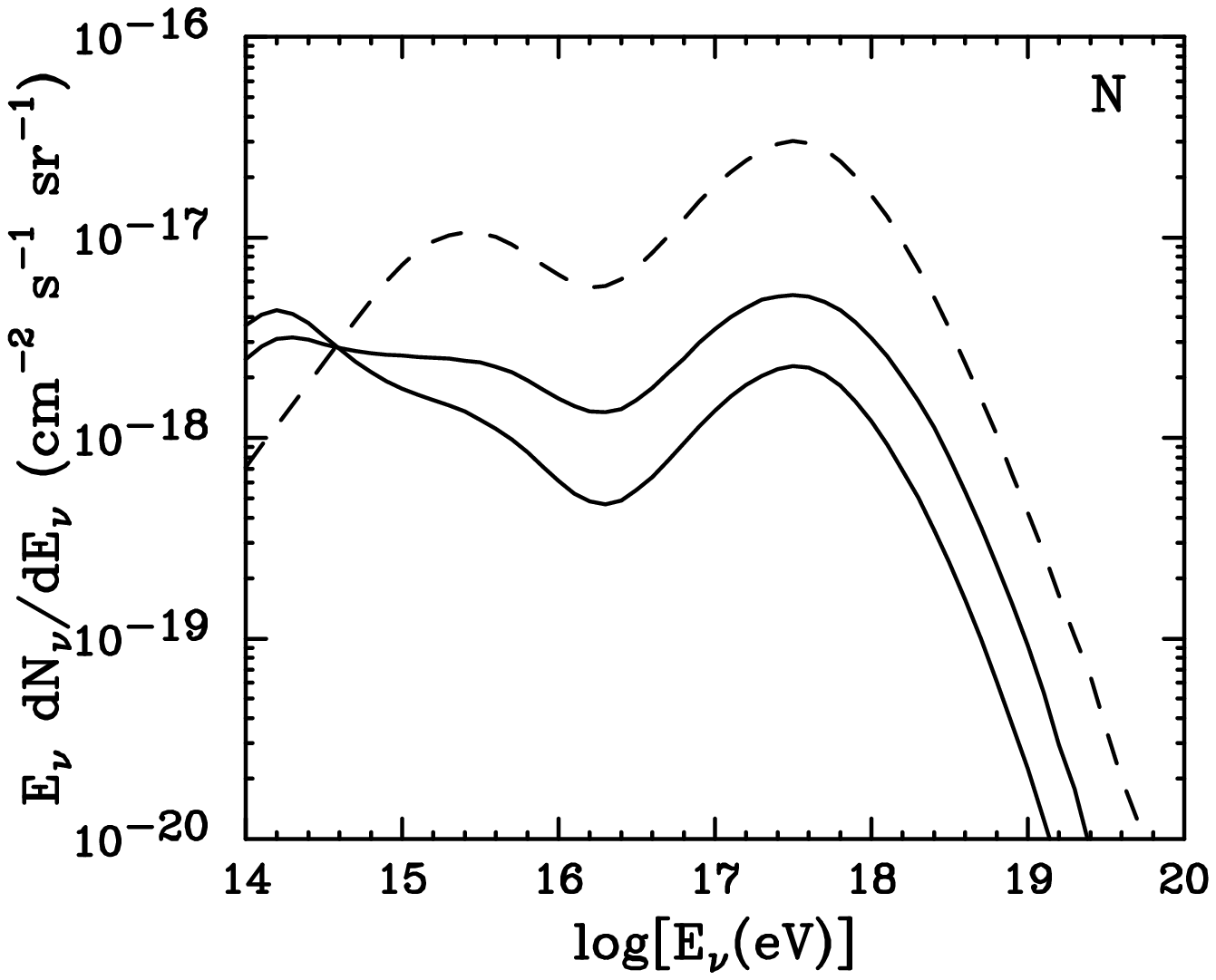,width=.35\columnwidth}
\hspace{1.0cm}
\epsfig{file=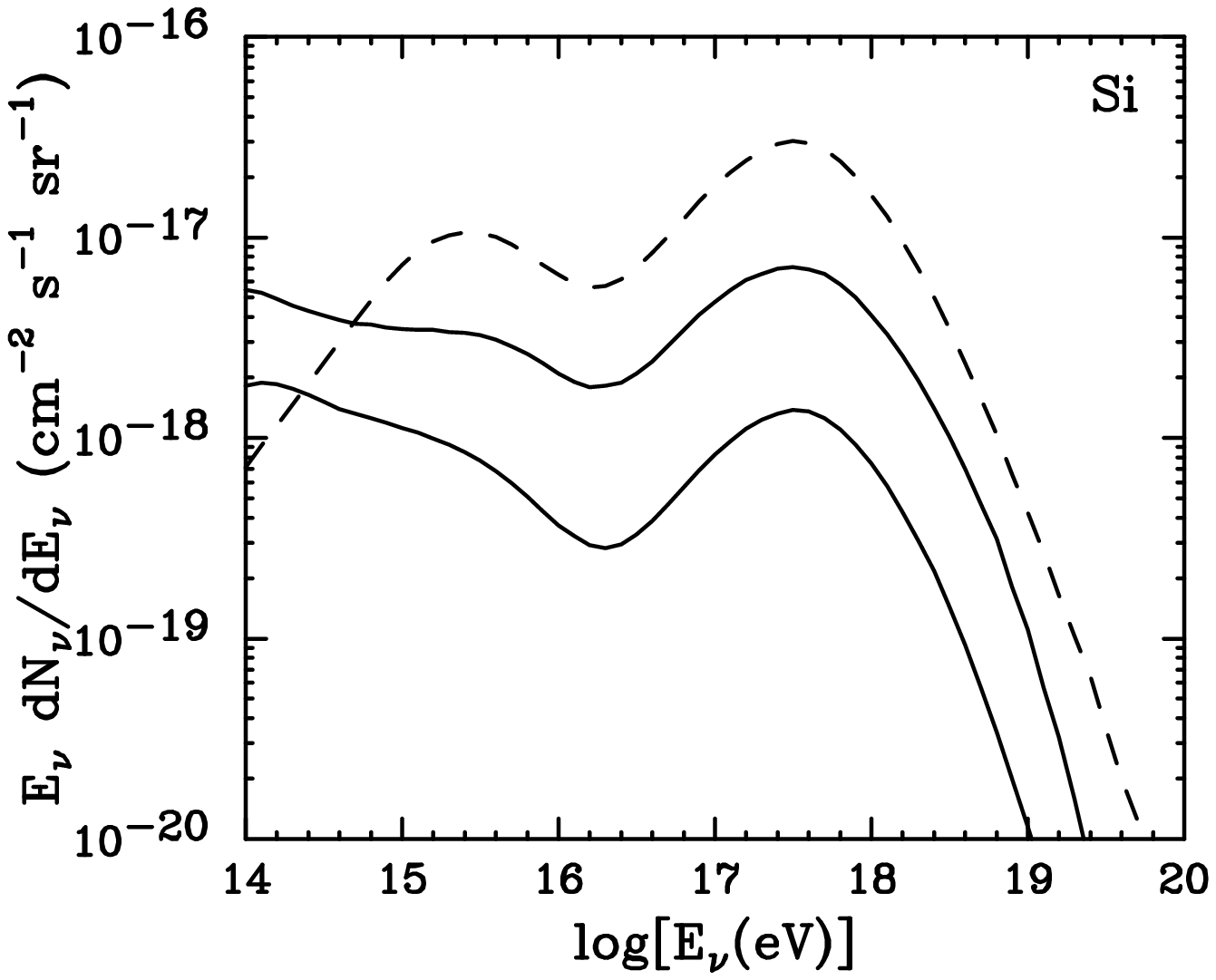,width=.35\columnwidth}\\ 
\epsfig{file=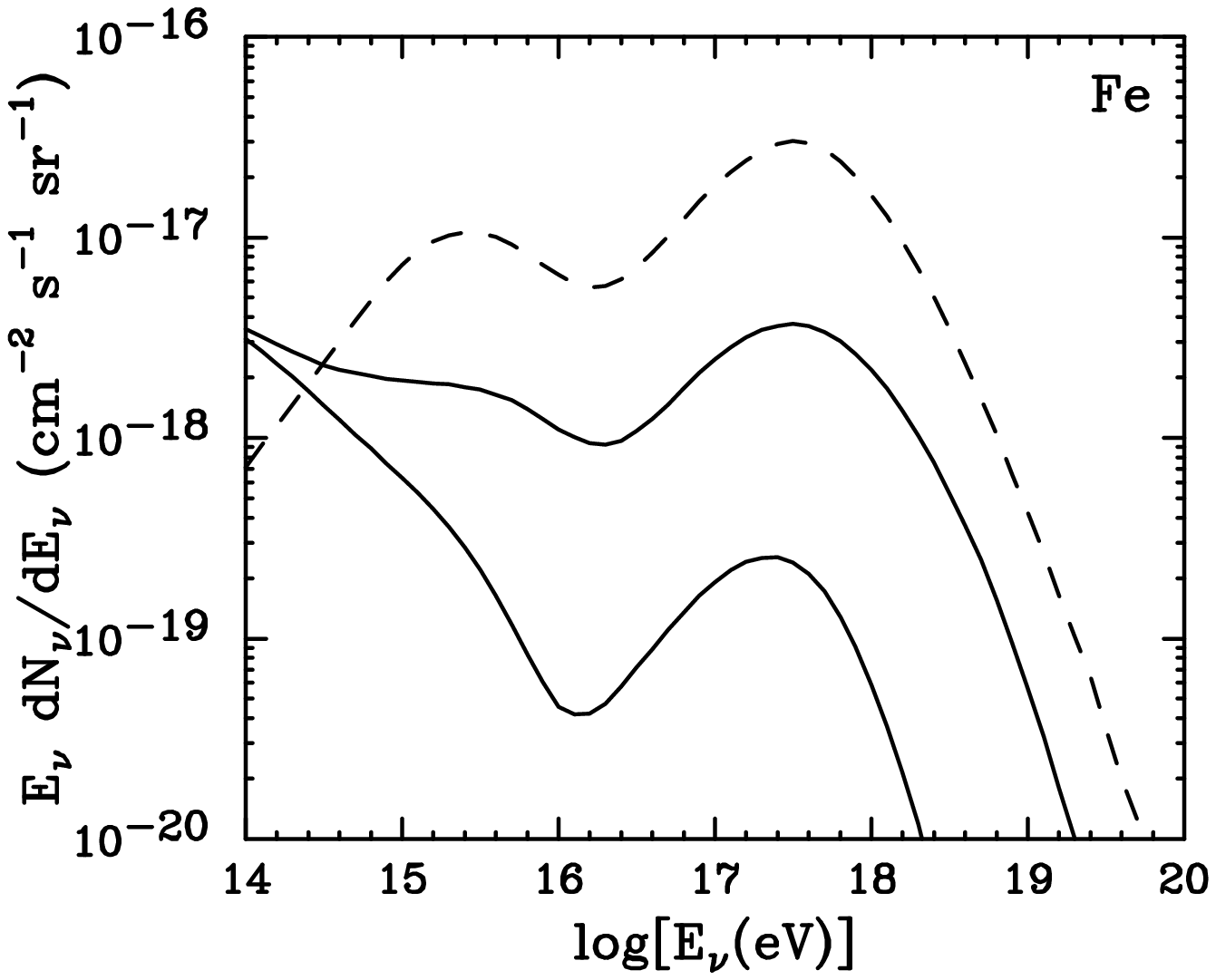,width=.35\columnwidth}
\hspace{1.0cm}
\epsfig{file=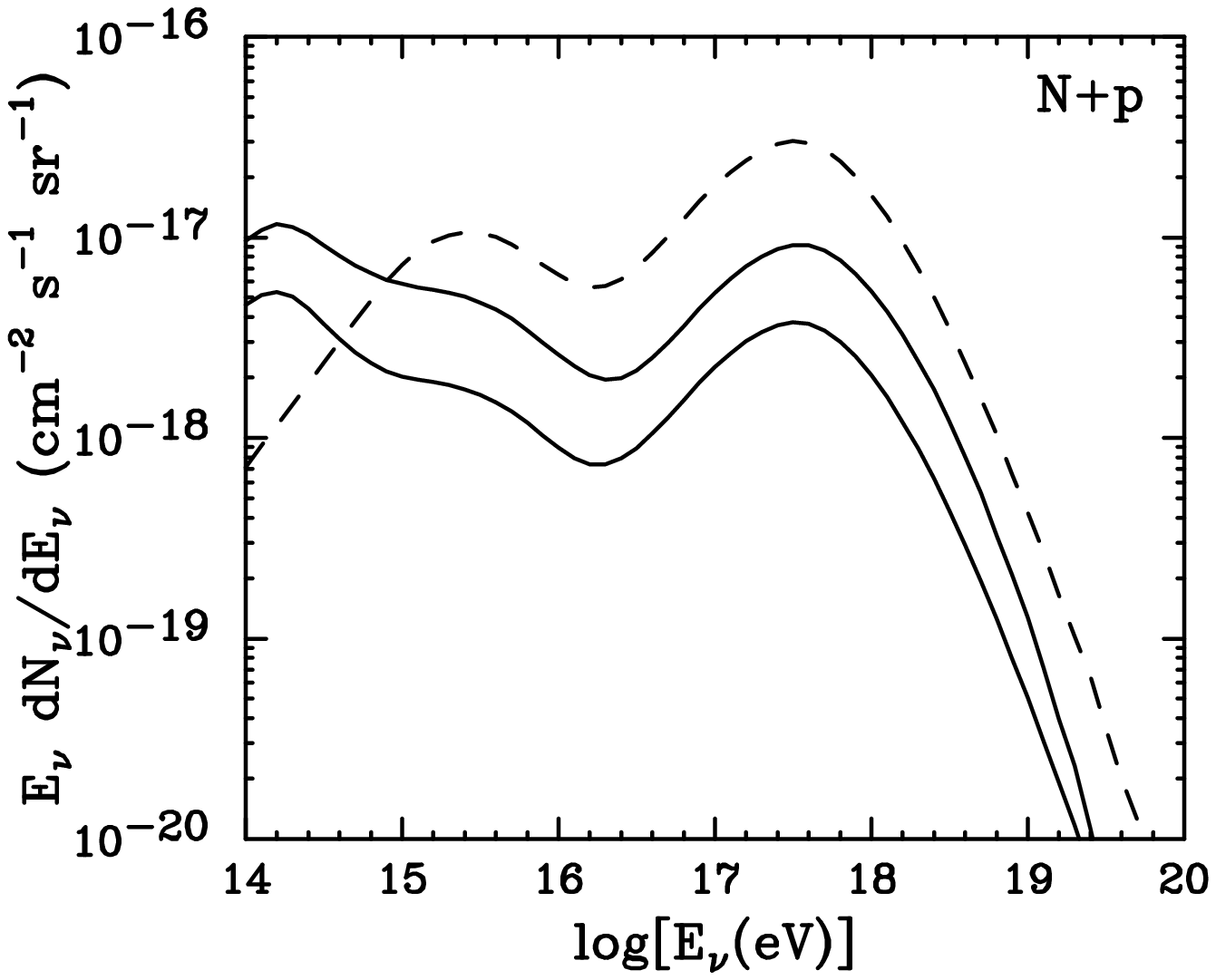,width=.35\columnwidth}\\
\epsfig{file=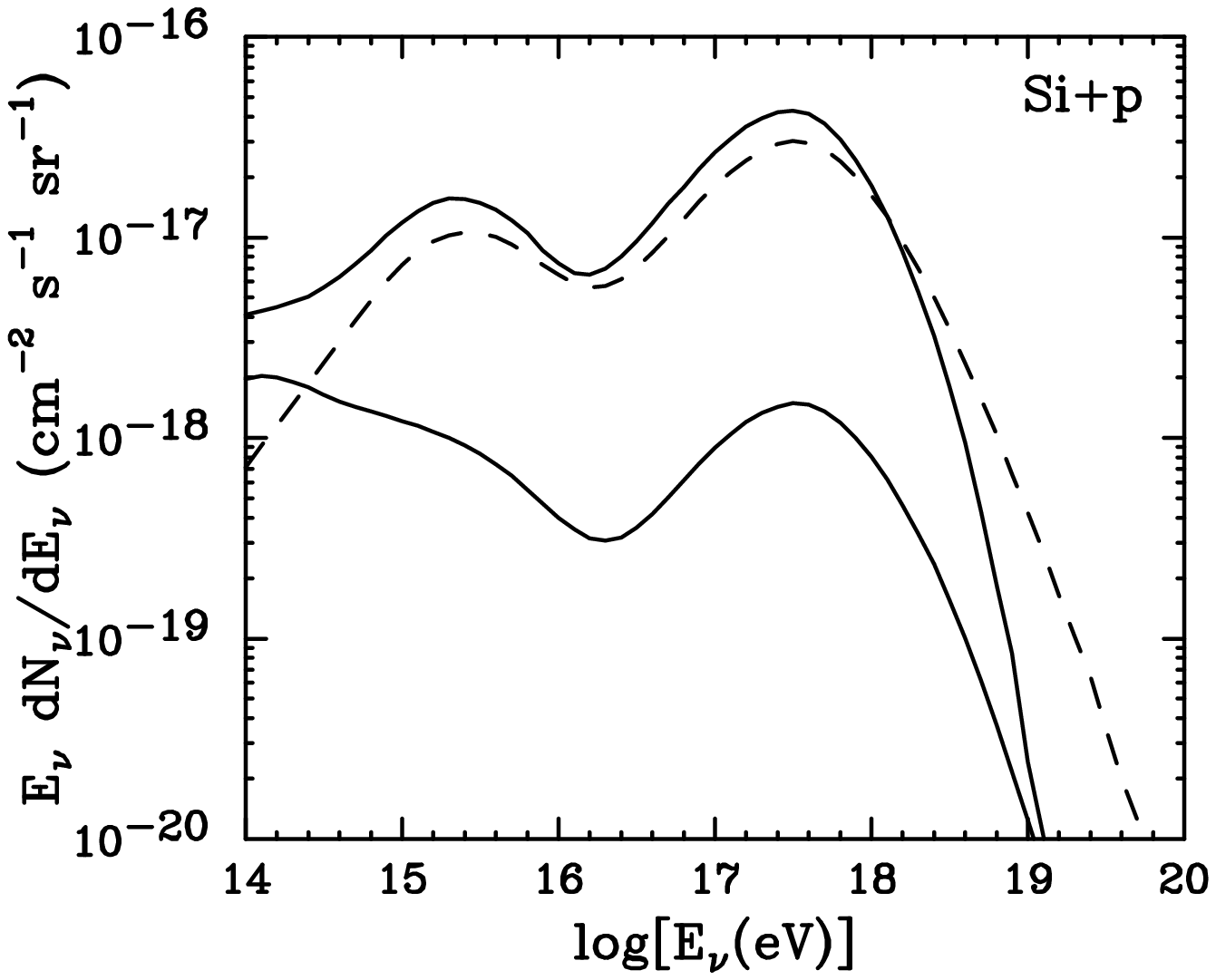,width=.35\columnwidth}
\hspace{1.0cm}
\epsfig{file=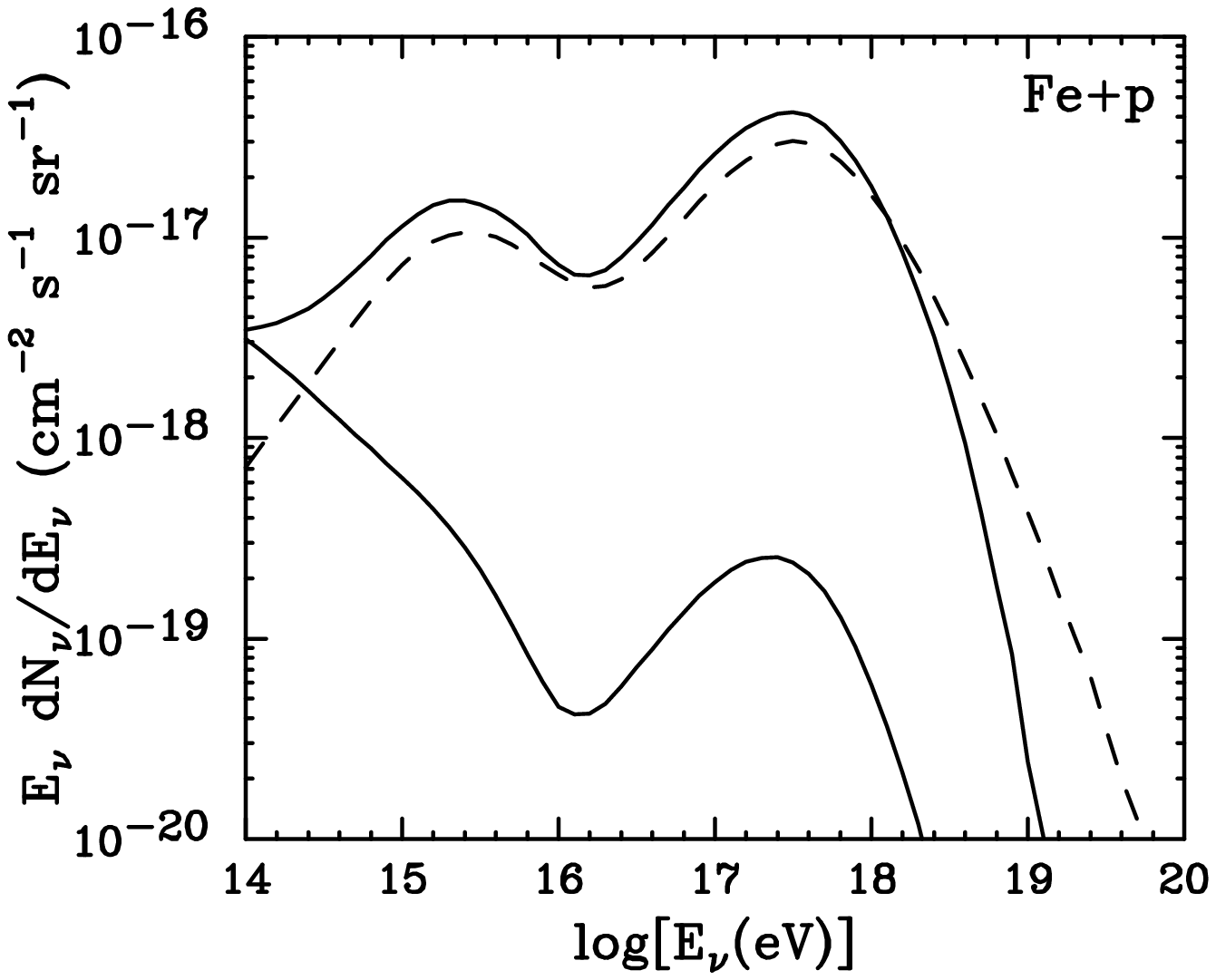,width=.35\columnwidth}\\
\caption{The range of cosmogenic neutrino spectra we find for various
  chemical species which are consistent with both the PAO spectrum and
  $X_{\rm{max}}$ measurements. In each case, we have considered model
  parameters in the range $\alpha=1.4-3.0$ and $E_{\max}/Z =
  10^{20}-10^{22}$ eV (although models with $E_{\max}/Z$ below
  approximately $10^{21}$ eV were found to be inconsistent with the
  data). In the N+p, Si+p and Fe+p frames, we show the results for
  combinations of injected nuclei and protons. In each frame, we show
  for comparison as a dashed curve the prediction for an all-proton
  spectrum with $\alpha=2.2$ and $E_{\rm{max}} =10^{22}$ eV. The solid
  lines denote the models with the highest and lowest rates predicted
  in a neutrino telescope such as IceCube.}
\label{neutrinospec}
\end{figure}

UHECR nuclei also interact with the cosmic microwave and infrared
backgrounds, undergoing photodisintegration. The disassociated
nucleons then interact with the cosmic microwave and infrared
backgrounds to produce cosmogenic neutrinos. In the limit that the
cosmic backgrounds are opaque to cosmic ray nuclei, full
disintegration occurs and the resulting cosmogenic neutrino spectrum
is not dramatically different from that predicted in the all-proton
case (assuming the cosmic ray spectrum extends to high enough energies
to produce protons above the GZK cutoff). In contrast, if a
significant fraction of cosmic ray nuclei remain intact, the resulting
flux of cosmogenic neutrinos can be considerably suppressed.

The predicted neutrino flux depends on the chemical composition and
spectrum of the injected cosmic rays. In Fig.~\ref{neutrinospec}, we plot the spectrum of the cosmogenic neutrinos for various scenarios. In each frame, we show the
maximal and minimal neutrino spectra (in terms of the resulting event
rate in a neutrino telescope) for a wide range of spectral parameters
($\alpha$, $E_{\rm{max}}$ and normalization) which were found to be
consistent with the PAO measurements of the UHECR spectrum and
elongation rate. We have considered values of these parameters in the
range of $\alpha=$ 1.4 to 3.0 and $E_{\rm{max}}/Z=10^{20}$ to
$10^{22}$ eV. In the first three frames, we have assumed pure
nitrogen, silicon and iron at injection, respectively, and find the
resulting neutrino flux to be suppressed compared to the all-proton
prediction by a factor between approximately 3 and 100. The smallest cosmogenic flux
is found in the case of an all-iron injected spectrum, exponentially
cutoff above $10^{21}$ eV. In this case, the spectrum of disassociated
protons is cutoff above $10^{21}$ eV/56 $\approx 2 \times 10^{19}$ eV,
which is slightly below the GZK cutoff.

In the last three frames of Fig.~\ref{neutrinospec}, we consider the
case of a mixed composition at injection. We find that the PAO data
can be fitted if 50--100\% of the injected particles are nitrogen
nuclei.  Mixtures of silicon and iron nuclei with protons, in
contrast, are acceptable with as little as 2\% and 1\% heavy nuclei,
respectively. In these cases the cosmogenic neutrino flux is only
slightly altered from the all-proton prediction.

A number of experimental programs are underway to detect high and
ultrahigh energy cosmic neutrinos (for reviews, see
Ref.~\cite{review}). These efforts include large volume detectors,
such as IceCube at the South Pole~\cite{Ahrens:2002dv} and KM3 in the
Mediterranean~\cite{km3}. In Table~\ref{t1}, we show the event
rates predicted in a kilometer-scale neutrino telescope for various
choices of injected spectra and chemical composition found to be
consistent with the PAO spectrum and elongation rate measurements. The
rates were calculated following the treatment described in
Ref.~\cite{icecubeplus}.

Although we have not calculated the corresponding rates here, other
experimental efforts to observe ultrahigh energy cosmic neutrinos
include the balloon-borne radio detector ANITA~\cite{Barwick:2005hn},
the under ice radio array RICE~\cite{rice}, and the PAO itself, which
hopes to detect neutrinos as either quasi-horizontal,
deeply-penetrating events, or as Earth-skimming, tau-neutrino induced
events~\cite{Bigas:2007tp}. Future programs to use underwater acoustic
detectors~\cite{acoustic}, radio antennas in rock
salt~\cite{rocksalt}, extensions of IceCube~\cite{icecubeplus} and
space-based cosmic ray detectors ({\it i.e.} JEM/EUSO)~\cite{euso} are
also being considered.

We have neglected in this study the effects of extragalactic magnetic fields on the propagation of UHECRs. If such particles are significantly deflected over cosmological distances, then the cosmogenic neutrino flux may be somewhat altered from the results shown here. We also note that event rates at neutrino telescopes
are expected to have a significant contribution from neutrino emission from the sources themselves. The impact of heavy nuclei in the injected
cosmic rays on such neutrino fluxes has been
discussed elsewhere~\cite{Anchordoqui:2007tn}.

\vspace{0.5cm}
 \begin{table}[!ht]
 \hspace{0.0cm}
 \begin{tabular} {c c c c c c c c c c c c c c c c c c c c c c} 
\hline \hline
 Composition &\vline& $\alpha$ & \vline &  $E_{\rm{max}}/Z$ &\vline&  Muons (km$^{-2}$ yr$^{-1}$) &\vline& Showers (km$^{-3}$ yr$^{-1}$)   \\
 \hline \hline
100\% N &\vline& 1.6-1.9 & \vline & $10^{22}$ eV & \vline & 0.20-0.0081 & \vline & 0.15-0.0064 \\
 \hline \hline
100\% Si &\vline& 1.6-2.0 & \vline & $10^{22}$ eV & \vline & 0.21-0.045 & \vline & 0.16-0.035 \\
 \hline \hline
100\% Fe &\vline& 1.6-2.1 & \vline & $10^{22}$ eV & \vline & 0.11-0.014 & \vline & 0.085-0.012 \\
\hline
100\% Fe &\vline& 1.4-1.7 & \vline & $10^{21}$ eV & \vline & 0.019-0.0076 & \vline & 0.017-0.0075 \\
\hline \hline
50\% N, 50\% p &\vline& 1.8-2.1 & \vline & $10^{22}$ eV & \vline & 0.23-0.13 & \vline & 0.18-0.10 \\
 \hline \hline
50\% Si, 50\% p &\vline& 1.6-2.1 & \vline & $10^{22}$ eV & \vline & 0.30-0.095 & \vline & 0.22-0.075 \\
 \hline
50\% Si, 50\% p &\vline& 1.4-1.5 & \vline & $10^{21}$ eV & \vline & 0.059-0.051 & \vline & 0.050-0.043 \\
 \hline 
7\% Si, 93\% p &\vline& 2.0-2.2 & \vline & $10^{22}$ eV & \vline & 0.69-0.66 & \vline & 0.52-0.50 \\
 \hline
2\% Si, 98\% p &\vline& 1.4-1.8 & \vline & $10^{21}$ eV & \vline & 0.75-0.59 & \vline & 0.60-0.47 \\
 \hline \hline
50\% Fe, 50\% p &\vline& 1.6-2.1 & \vline & $10^{22}$ eV & \vline & 0.15-0.043 & \vline & 0.11-0.034 \\
 \hline
10\% Fe, 90\% p &\vline& 1.4-1.9 & \vline & $10^{21}$ eV & \vline & 0.14-0.10 & \vline & 0.11-0.080 \\
 \hline
3\% Fe, 97\% p &\vline& 2.1 & \vline & $10^{22}$ eV & \vline & 0.68 & \vline & 0.51 \\
 \hline
1\% Fe, 99\% p &\vline& 1.4-1.9 & \vline & $10^{21}$ eV & \vline & 0.74-0.53 & \vline & 0.59-0.43 \\
 \hline \hline
100\% p (for comparison)  &\vline& 2.2 & \vline & $10^{22}$ eV & \vline & 0.76 & \vline & 0.60 \\
\hline \hline
 \end{tabular}
 \caption{The rates of muon and shower events in a kilometer-scale neutrino telescope (such as IceCube or KM3) from cosmogenic neutrinos for a range of choices of injected spectra and chemical composition consistent with both the PAO spectrum and $X_{\rm{max}}$ measurements. For comparison, we have also shown the event rates for the case of an all-proton spectrum with $\alpha=2.2$ and $E_{\rm{max}} =10^{22}$ eV. We find models consistent with the PAO data which predict rates very similar to those found in the all-proton case, and models in which the event rates are suppressed by up to two order of magnitude.}
\label{t1}
 \end{table}

\section{Summary And Conclusions}

Utilizing as inputs the energy spectrum and composition of the
ultrahigh energy cosmic rays injected from extragalactic
sources, we have determined a range of models consistent with the
recent PAO measurements of the cosmic ray spectrum and elongation
rates. The results of our analysis may be summarized as follows:

\begin{itemize}
\item{Consistency with the data can be achieved if the injected
    spectrum consists of either $(a)$ entirely intermediate mass (C,
    N, O) or heavy (Si, Fe) nuclei, or $(b)$ protons with even a very
    small admixture (1-10\%) of heavy elements. An injected
    composition consisting solely of protons or light nuclei, in
    contrast, does not provide an acceptable fit to the elongation
    rate data.}


\item{If the UHECR spectrum is dominated by intermediate mass or heavy
    nuclei, the resulting neutrino flux (from the photopion interactions of
    nucleons liberated through photodisintegration, or via neutron
    decay) is expected to be suppressed relative to the all-proton
    prediction by a factor of approximately 3 to 100, leading to event rates at
    IceCube within the range of approximately 0.2-0.007 muons and
    0.16-0.006 showers per year. This suppression is, in part, the result of the
    disassociated nucleons possessing too little energy to undergo
    photopion interactions with CMB photons ({\it i.e.} they are below
    the ``GZK cutoff''). An injected composition of protons with a
    very small (1-10\%) admixture of heavy nuclei, however, will
    generate a cosmogenic neutrino flux similar to an all-proton
    input, while simultaneously providing a fully acceptable fit to
    the PAO data.}
\end{itemize}
As a final comment, we note that if future observations reveal an
anisotropic distribution of UHECR arrival directions (such as point
sources, for example), then a significant fraction of the cosmic rays are likely to
be protons rather than heavy nuclei (whose large electric charge would
lead to a loss of directionality in the intervening magnetic fields).
Since an injected composition consisting solely of protons is
disallowed by the PAO data, the only remaining possibility would be a
proton-dominated injection spectrum with a small admixture of heavy
nuclei. This would imply that neutrino telescopes such as IceCube will attain discovery reach
of a high energy diffuse neutrino flux within a few years of
observations.

\acknowledgements{We would like to thank Tom Weiler, Francis Halzen
  and Teresa Montaruli for interesting comments and discussion. This
  work has been supported by the US Department of Energy, the National
  Science Foundation grant PHY-0224507, NASA grant NAG5-10842, STFC Senior Fellowship
  PPA/C506205/1, a MPIK Fellowship, and the EU network `UniverseNet'
  (MRTN-CT-2006-035863).}


\begin{thebibliography}{99}

\bibitem{Abraham:2004dt}
  J.~Abraham {\it et al.}  [Pierre Auger Collaboration],
  Nucl.\ Instrum.\ Meth.\ A {\bf 523}, 50 (2004).

\bibitem{hillas}
A.~M.~Hillas,
Ann.\ Rev.\ Astron.\ Astrophys.\ {\bf 22} (1984) 425.


\bibitem{berezinsky}
  V.~Berezinsky, A.~Z.~Gazizov and S.~I.~Grigorieva,
  Phys.\ Lett.\  B {\bf 612}, 147 (2005)
  [arXiv:astro-ph/0502550];
  R.~Aloisio, V.~Berezinsky, P.~Blasi, A.~Gazizov, S.~Grigorieva and B.~Hnatyk,
  Astropart.\ Phys.\  {\bf 27}, 76 (2007)
  [arXiv:astro-ph/0608219];
  R.~Aloisio, V.~Berezinsky, P.~Blasi and S.~Ostapchenko,
  arXiv:0706.2834 [astro-ph].




\bibitem{Anchordoqui:2004xb} 
  L.~Anchordoqui, M.~T.~Dova, A.~Mariazzi, T.~McCauley, T.~Paul, 
  S.~Reucroft and J.~Swain,
  Annals Phys.\  {\bf 314}, 145 (2004)
  [arXiv:hep-ph/0407020].

\bibitem{Linsley:gh}
  J.~Linsley and A.~A.~Watson,
  Phys.\ Rev.\ Lett.\  {\bf 46}, 459 (1981); and references therein.



\bibitem{Roth:2007in}
  M.~Roth {\it et al.} [Pierre Auger Collaboration],
  arXiv:0706.2096 [astro-ph];
  L.~Perrone {\it et al.} [Pierre Auger Collaboration],
  arXiv:0706.2643 [astro-ph];
  P.~Facal San Luis {\it et al.} [Pierre Auger Collaboration],
  arXiv:0706.4322 [astro-ph].


\bibitem{Unger:2007mc}
  M.~Unger {\it et al.} [Pierre Auger Collaboration],
  arXiv:0706.1495 [astro-ph].


\bibitem{Anchordoqui:2007tb}
  L.~Anchordoqui {\it et al.} [Pierre Auger Collaboration],
  arXiv:0706.0989 [astro-ph];
  D.~Harari {\it et al.} [Pierre Auger Collaboration],
  arXiv:0706.1715 [astro-ph];
  S.~Mollerach {\it et al.} [Pierre Auger Collaboration],
  arXiv:0706.1749 [astro-ph];
  E.~Armengaud {\it et al.} [Pierre Auger Collaboration],
  arXiv:0706.2640 [astro-ph].



\bibitem{Bigas:2007tp}
  O.~B.~Bigas  {\it et al.} [Pierre Auger Collaboration],
  arXiv:0706.1658 [astro-ph].



\bibitem{Abbasi:2004nz}
  R.~U.~Abbasi {\it et al.}  [HiRes Collaboration],
  Astrophys.\ J.\  {\bf 622}, 910 (2005)
  [arXiv:astro-ph/0407622].


\bibitem{gzk}
  K.~Greisen,
  Phys.\ Rev.\ Lett.\  {\bf 16}, 748 (1966);
  G.~T.~Zatsepin and V.~A.~Kuzmin,
  JETP Lett.\  {\bf 4}, 78 (1966)
  [Pisma Zh.\ Eksp.\ Teor.\ Fiz.\  {\bf 4}, 114 (1966)].


\bibitem{cosmogenic}
  V.~S.~Berezinsky and G.~T.~Zatsepin,
  Phys.\ Lett.\ B {\bf 28} 423 (1969);
  Yad.\ Fiz.\  {\bf 11}, 200 (1970);
  F.~W.~Stecker,
  Astrophys.\ J.\  {\bf 228}, 919 (1979);
  C.~T.~Hill and D.~N.~Schramm,
  Phys.\ Lett.\  B {\bf 131}, 247 (1983);
  R.~Engel, D.~Seckel and T.~Stanev,
  Phys.\ Rev.\  D {\bf 64}, 093010 (2001)
  [arXiv:astro-ph/0101216];
  Z.~Fodor, S.~D.~Katz, A.~Ringwald and H.~Tu,
  JCAP {\bf 0311}, 015 (2003)
  [arXiv:hep-ph/0309171].





\bibitem{Ahrens:2002dv}
  J.~Ahrens {\it et al.}  [The IceCube Collaboration],
  Nucl.\ Phys.\ Proc.\ Suppl.\  {\bf 118}, 388 (2003)
  [arXiv:astro-ph/0209556].

\bibitem{Barwick:2005hn}
  S.~W.~Barwick {\it et al.}  [ANITA Collaboration],
  Phys.\ Rev.\ Lett.\  {\bf 96}, 171101 (2006)
  [arXiv:astro-ph/0512265];
  P.~Miocinovic {\it et al.}  [The ANITA Collaboration],
  eConf {\bf C041213}, 2516 (2004)
  [arXiv:astro-ph/0503304].
 
\bibitem{augernus}
  K.~S.~Capelle, J.~W.~Cronin, G.~Parente and E.~Zas,
  Astropart.\ Phys.\  {\bf 8}, 321 (1998)
  [arXiv:astro-ph/9801313];
  D.~Fargion,
  Astrophys.\ J.\  {\bf 570}, 909 (2002)
  [arXiv:astro-ph/0002453];
  X.~Bertou, P.~Billoir, O.~Deligny, C.~Lachaud and A.~Letessier-Selvon,
  Astropart.\ Phys.\  {\bf 17}, 183 (2002)
  [arXiv:astro-ph/0104452];
  J.~L.~Feng, P.~Fisher, F.~Wilczek and T.~M.~Yu,
  Phys.\ Rev.\ Lett.\  {\bf 88}, 161102 (2002)
  [arXiv:hep-ph/0105067];
  S.~Palomares-Ruiz, A.~Irimia and T.~J.~Weiler,
  Phys.\ Rev.\  D {\bf 73}, 083003 (2006)
  [arXiv:astro-ph/0512231].


\bibitem{cosmogenicnuclei}
  D.~Hooper, A.~Taylor and S.~Sarkar,
  Astropart.\ Phys.\  {\bf 23}, 11 (2005)
  [arXiv:astro-ph/0407618];
  M.~Ave, N.~Busca, A.~V.~Olinto, A.~A.~Watson and T.~Yamamoto,
  Astropart.\ Phys.\  {\bf 23}, 19 (2005)
  [arXiv:astro-ph/0409316];
  D.~Allard {\it et al.},
  JCAP {\bf 0609}, 005 (2006)
  [arXiv:astro-ph/0605327].



\bibitem{Hillas:1985is} 
  A.~M.~Hillas,
  Ann.\ Rev.\ Astron.\ Astrophys.\  {\bf 22}, 425 (1984).


\bibitem{Werner:2007vd}
  K.~Werner and T.~Pierog,
  arXiv:0707.3330 [astro-ph].

\bibitem{Kalmykov:1997te}
  N.~N.~Kalmykov, S.~S.~Ostapchenko and A.~I.~Pavlov,
  Nucl.\ Phys.\ Proc.\ Suppl.\  {\bf 52B}, 17 (1997).

\bibitem{Fletcher:1994bd}
  R.~S.~Fletcher, T.~K.~Gaisser, P.~Lipari and T.~Stanev,
  Phys.\ Rev.\ D {\bf 50}, 5710 (1994);
  J.~Engel, T.~K.~Gaisser, T.~Stanev and P.~Lipari,
  Phys.\ Rev.\ D {\bf 46}, 5013 (1992).

\bibitem{Hooper:2006tn}
  D.~Hooper, S.~Sarkar and A.~M.~Taylor,
  Astropart.\ Phys.\  {\bf 27}, 199 (2007)
  [arXiv:astro-ph/0608085].

\bibitem{other}
  F.~W.~Stecker and M.~H.~Salamon,
  Astrophys.\ J.\  {\bf 512}, 521 (1999)
  [arXiv:astro-ph/9808110];
  L.~N.~Epele and E.~Roulet,
  JHEP {\bf 9810}, 009 (1998)
  [arXiv:astro-ph/9808104];
  E.~Armengaud, G.~Sigl and F.~Miniati,
  Phys.\ Rev.\  D {\bf 72}, 043009 (2005)
  [arXiv:astro-ph/0412525];
  G.~Sigl and E.~Armengaud,
  JCAP {\bf 0510}, 016 (2005)
  [arXiv:astro-ph/0507656];
  G.~Bertone, C.~Isola, M.~Lemoine and G.~Sigl,
  Phys.\ Rev.\  D {\bf 66}, 103003 (2002)
  [arXiv:astro-ph/0209192];
  T.~Yamamoto, K.~Mase, M.~Takeda, N.~Sakaki and M.~Teshima,
  Astropart.\ Phys.\  {\bf 20}, 405 (2004)
  [arXiv:astro-ph/0312275];
  E.~Khan {\it et al.},
  Astropart.\ Phys.\  {\bf 23}, 191 (2005)
  [arXiv:astro-ph/0412109];
  D.~Allard, E.~Parizot, E.~Khan, S.~Goriely and A.~V.~Olinto,
  Astron.\ Astrophys.\  {\bf 443}, L29 (2005)
  [arXiv:astro-ph/0505566].





\bibitem{Bird:1993yi}
  D.~J.~Bird {\it et al.}  [HIRES Collaboration],
  Phys.\ Rev.\ Lett.\  {\bf 71}, 3401 (1993).



\bibitem{yakutsk}
B.~N.~Afanasiev {\it et al.}  [Yakutsk Collaboration],
Proc.~Tokyo Workshop on Techniques for the Study of the Extremely High Energy Cosmic Rays, Ed. M.~Nagano (1993).



\bibitem{review}
  F.~Halzen and D.~Hooper,
  Rept.\ Prog.\ Phys.\  {\bf 65}, 1025 (2002)
  [arXiv:astro-ph/0204527];
 J.~G.~Learned and K.~Mannheim,
  Ann.\ Rev.\ Nucl.\ Part.\ Sci.\  {\bf 50}, 679 (2000).








\bibitem{km3}
  P.~Sapienza,
  Nucl.\ Phys.\ Proc.\ Suppl.\  {\bf 145}, 331 (2005).

\bibitem{icecubeplus} 
  F.~Halzen and D.~Hooper,
  JCAP {\bf 0401}, 002 (2004)
  [arXiv:astro-ph/0310152].




\bibitem{rice}
 I.~Kravchenko {\it et al.},
  arXiv:astro-ph/0601148.
 




\bibitem{acoustic}
 N.~G.~Lehtinen, S.~Adam, G.~Gratta, T.~K.~Berger and M.~J.~Buckingham,
  Astropart.\ Phys.\  {\bf 17}, 279 (2002)
  [arXiv:astro-ph/0104033];
  L.~G.~Dedenko, I.~M.~Zheleznykh, S.~K.~Karaevsky, A.~A.~Mironovich, V.~D.~Svet and A.~V.~Furduev,
  Bull.\ Russ.\ Acad.\ Sci.\ Phys.\  {\bf 61}, 469 (1997)
  [Izv.\ Ross.\ Akad.\ Nauk.\  {\bf 61}, 593 (1997)];
 D.~Besson, S.~Boser, R.~Nahnhauer, P.~B.~Price and J.~A.~Vandenbroucke
                  [IceCube Collaboration],
  arXiv:astro-ph/0512604.


\bibitem{rocksalt}
  P.~W.~Gorham, D.~Saltzberg, R.~C.~Field, E.~Guillian, R.~Milincic, 
  D.~Walz and D.~Williams,
  Phys.\ Rev.\ D {\bf 72}, 023002 (2005)
  [arXiv:astro-ph/0412128];
  P.~Gorham, D.~Saltzberg, A.~Odian, D.~Williams, D.~Besson, G.~Frichter and S.~Tantawi,
  Nucl.\ Instrum.\ Meth.\ A {\bf 490}, 476 (2002)
  [arXiv:hep-ex/0108027].


\bibitem{euso}
http://euso.riken.go.jp/




\bibitem{Anchordoqui:2007tn}
  L.~A.~Anchordoqui, D.~Hooper, S.~Sarkar and A.~M.~Taylor,
  arXiv:astro-ph/0703001.






\end{thebibliography}
\end{document}